\documentclass[conference]{IEEEtran}
\makeatletter
\def\ps@headings{%
\def\@oddhead{\mbox{}\scriptsize\rightmark \hfil \thepage}%
\def\@evenhead{\scriptsize\thepage \hfil \leftmark\mbox{}}%
\def\@oddfoot{}%
\def\@evenfoot{}}
\makeatother
\usepackage[pdftex]{graphicx}
\usepackage{amsmath}
\usepackage{amssymb}
\usepackage{array}
\usepackage[svgnames]{xcolor} 
\usepackage{mdwmath}
\usepackage{subfigure}
 \usepackage{bm}
\usepackage{tipa}
 \usepackage{hyperref}
 \usepackage{framed}
 \usepackage{booktabs}
 \usepackage{caption}

\usepackage{algorithm}
\usepackage{algorithmic}
\usepackage{epsfig}
\usepackage{overpic}
\usepackage{epstopdf}
\usepackage{amsfonts,setspace,amsthm,braket}

\graphicspath{{../pdf/}{../jpeg/}}
\DeclareGraphicsExtensions{.pdf,.jpeg,.png}

\newtheorem{theorem}{Theorem}
\newtheorem{lemma}{Lemma}

\newtheorem{definition}{Definition}

\newtheorem{constraint}{Constraint}

\allowdisplaybreaks

\newcommand{\conv}[1]{\operatorname{conv}\left(#1\right)}

\newcommand{\beqa}{\begin{eqnarray}}
\newcommand{\eeqa}{\end{eqnarray}}

\floatname{algorithm}{Algorithm}

\begin{document}
\title{Throughput-Optimal Broadcast on \\ Directed Acyclic Graphs} 

\author{
\IEEEauthorblockN{Abhishek Sinha, Georgios Paschos, 
Chih-ping Li, and Eytan Modiano}
\IEEEauthorblockA{Laboratory for Information and Decision Systems, Massachusetts Institute of Technology, Cambridge, MA 02139\\
Email: \{sinhaa, gpasxos, cpli, modiano\}@mit.edu}
}
\maketitle
\begin{abstract}
We study the problem of broadcasting packets in wireless networks.
At each time slot, a network controller activates non-interfering links and forwards packets to all nodes at a common rate; the maximum rate is referred to as the broadcast capacity of the wireless network.
Existing policies achieve the broadcast capacity 
by  balancing traffic over a set of spanning trees, 
which are difficult to maintain in a large and time-varying wireless network.
We propose a new dynamic algorithm that achieves the broadcast capacity when the underlying network topology is a directed acyclic graph (DAG). This algorithm utilizes local queue-length information, does not use any global topological structures such as spanning trees, and uses the idea of in-order packet delivery to all network nodes.
Although the in-order packet delivery constraint leads to degraded throughput in cyclic graphs, we show that it is throughput optimal in DAGs and can be exploited to simplify the design and analysis of optimal algorithms. 
Our simulation results show that the proposed algorithm has superior delay performance as compared to tree-based approaches.
\end{abstract}

\section{Introduction}
Broadcast refers to the fundamental network functionality of delivering data from a source node to all other nodes. 
It uses packet replication and appropriate forwarding to eliminate unnecessary packet retransmissions.
This is especially important in power-constrained wireless systems which suffer from interference and collisions.
Broadcast applications include mission-critical military communications \cite{military},  
live video streaming \cite{lstream}, and data dissemination in sensor networks \cite{akyildiz2002}.

The design of efficient wireless broadcast algorithms faces several challenges.
Wireless channels suffer from  interference, and a broadcast policy needs to activate non-interfering links at any time. Wireless network topologies undergo frequent changes, so that packet forwarding decisions must be made in an adaptive fashion. Existing dynamic multicast algorithms that balance traffic over spanning trees~\cite{swati} may be used for broadcasting, since broadcast is a special case of multicast.
These algorithms, however, are not suitable for wireless networks because enumerating all spanning trees is computationally complex and needs to be performed repeatedly when the network topology changes.

In this paper, we study the fundamental performance of broadcasting packets in wireless networks.
We consider a time-slotted system. At every slot, a scheduler decides which wireless links to activate and which packets to forward on activated links, so that all nodes receive packets at a common rate. The broadcast capacity is the maximum common reception rate of distinct packets over all scheduling policies.
We then design optimal wireless broadcast algorithms without the use of spanning trees. To the best of our knowledge, there exists no capacity-achieving scheduling policy for wireless broadcast without spanning trees.

We begin by considering a rich class of scheduling policies $\Pi$ that perform arbitrary link activations and packet forwarding, and characterize the broadcast capacity over this policy class $\Pi$. We impose two additional constraints that improve the understanding of the problem. First, we consider the subclass of policies $\Pi^{\text{in-order}}\subset \Pi$ that enforce the in-order delivery of packets. Second, we focus on the subset of policies $\Pi^* \subset \Pi^{\text{in-order}}$ that allows the reception of a packet by a node only if all its  incoming neighbors have received the packet.
It is intuitively clear that the policies in the more structured class $\Pi^*$ are easier to describe and analyze, but may yield degraded throughput performance. We show the surprising result that when the underlying network topology is a directed acyclic graph (DAG), there is a control policy $\pi^* \in \Pi^*$ that achieves the broadcast capacity. In contrast, there exists a cyclic network in which no control policy in $\Pi^*$ or $\Pi^{\text{in-order}}$ can achieve the broadcast capacity.
 
To enable the design of the optimal broadcast policy, we establish a \emph{queue-like dynamics} for the system state, which is represented by relative packet deficit. This is non-trivial for the broadcast problem because explicit queueing structure is difficult to maintain in the network due to packet replication. As a result, achieving the broadcast capacity reduces to finding a scheduling policy that stabilizes the system states using the drift analysis~\cite{tassiulas,neely2010stochastic}.  

In this paper, our contributions include:
\begin{itemize}
\item We define the broadcast capacity of a wireless network and show that it is characterized by an edge-capacitated graph $\widehat {\cal G}$ that arises from optimizing the time-averages of link activations. For integral-capacitated DAGs, the broadcast capacity is determined by the minimum in-degree of the graph $\widehat {\cal G}$, which is equal to the maximal number of edge-disjoint spanning trees.
\item We design a dynamic algorithm that utilizes local queue-length information to achieve the broadcast capacity of a wireless DAG network. This algorithm does not rely on spanning trees, has small computational complexity, and is suitable for mobile networks with time-varying topology. 
\item 
We demonstrate the superior delay performance of our algorithm, as compared to centralized tree-based algorithm \cite{swati}, via numerical simulations.
\end{itemize}
 
In the literature, a simple method for wireless broadcast is to use packet flooding~\cite{sasson2003probabilistic}.
The flooding approach, however, leads to redundant transmissions and collisions, known as \emph{broadcast storm} \cite{tseng2002broadcast}.
In the wired domain, it has been shown that forwarding \emph{useful} packets at random is optimal for broadcast \cite{massoulie2007randomized}; this approach does not extend to the wireless setting due to interference and the need for scheduling~\cite{Towsley2008}. Broadcast on wired networks can also be done using network coding~\cite{rate,Ho2005}. However, efficient link activation under network coding remains an open problem.


The rest of the paper is organized as follows. 
Section \ref{System model1} introduces the wireless network model. In Section \ref{capacity_section}, we define the broadcast capacity of a wireless network and provide a useful upper bound from a fundamental cut-set bound. In Section \ref{sec:algorithm}, we propose a dynamic broadcast policy that achieves the broadcast capacity in a DAG. Simulation results are presented in Section \ref{sec:simulations}.

\section{The Wireless Network Model} \label{System model1}

\begin{figure}[h!]
\centering
\subfigure[a wireless network]{
	\begin{overpic}[width=0.22\textwidth]{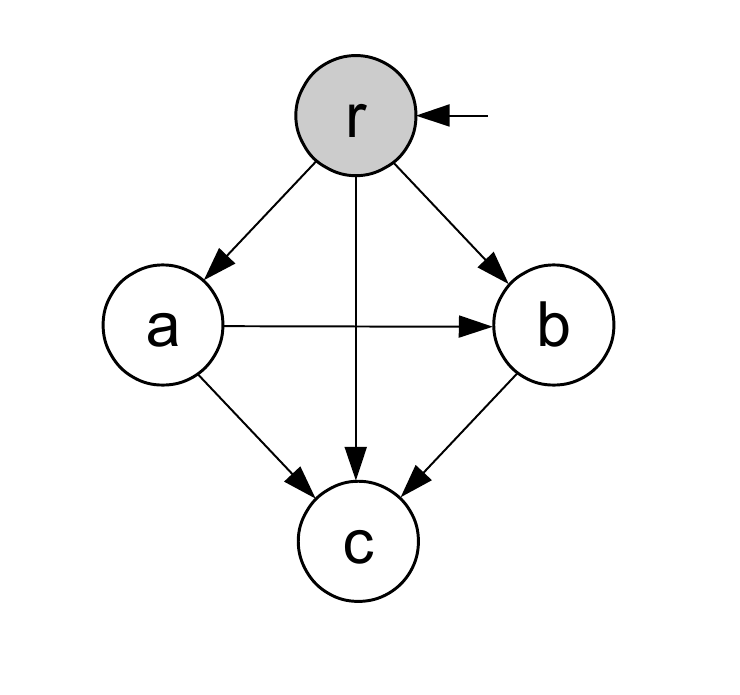}
	\put(69,77){$\lambda$}
	\end{overpic}
	\label{network-a}
}
\subfigure[activation vector $\bm{s}_1$]{
	\begin{overpic}[width=0.22\textwidth]{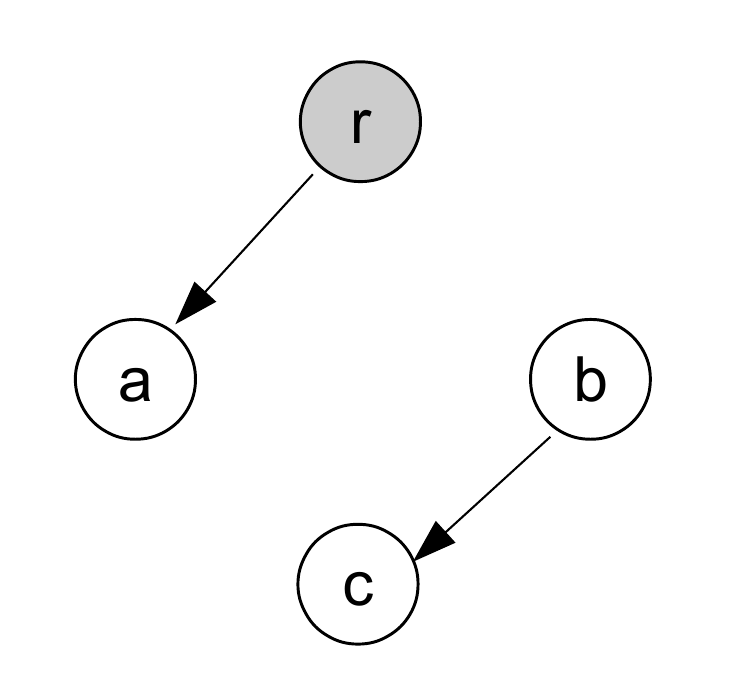}
	\end{overpic}
}
\subfigure[activation vector $\bm{s}_2$]{
  \begin{overpic}[width=0.22\textwidth]{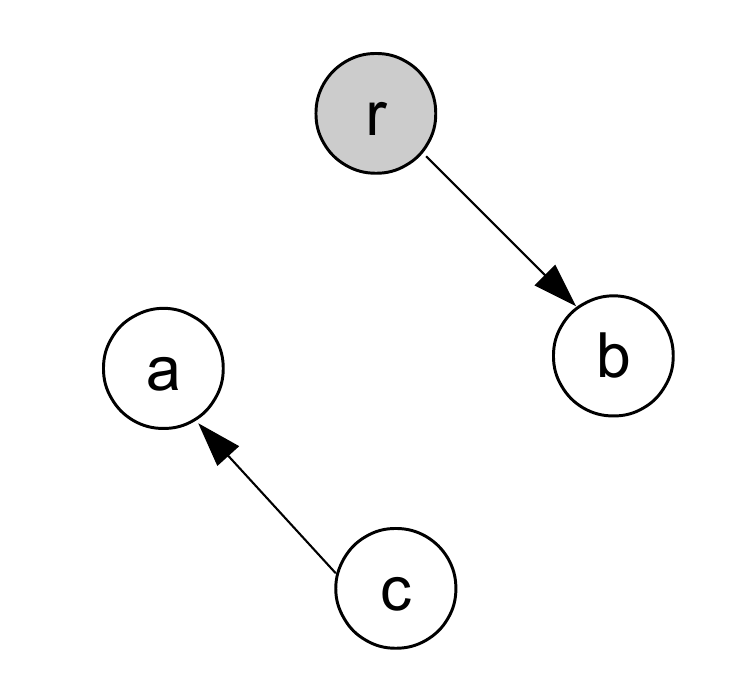}
  \end{overpic}
}
\subfigure[activation vector $\bm{s}_3$]{
  \begin{overpic}[width=0.22\textwidth]{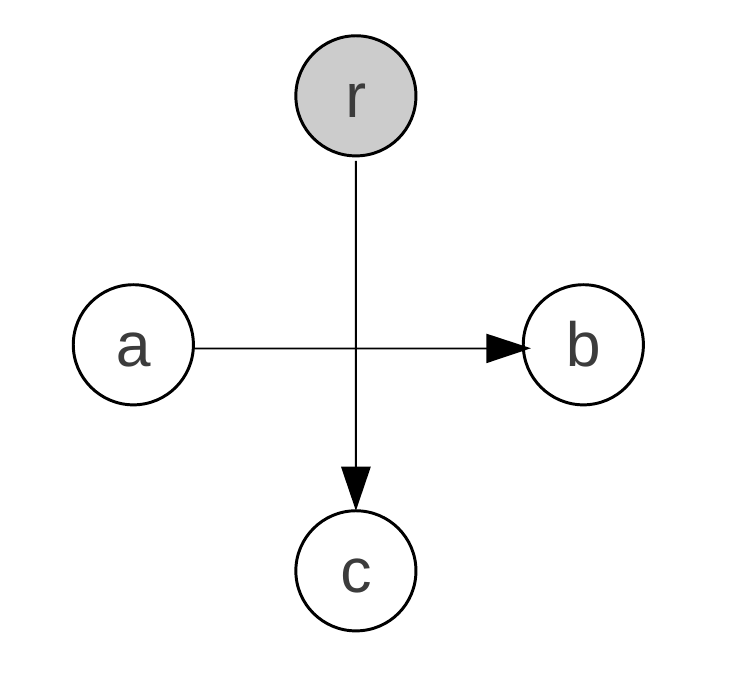}
  \end{overpic}
}
\caption{A wireless network and its three feasible link activations under the primary interference constraint.}
\label{network}
\end{figure}

We consider a wireless network that is represented by a directed graph $\mathcal{G}=(V,E,\bm{c},\mathcal{S})$, where $V$ is the set of nodes, $E$ is the set of directed links, $\bm{c} = (c_{ij})$ denotes the capacity of links $(i, j)\in E$, $\mathcal{S}$ is the set of all feasible link-activation vectors, and $\bm{s} = (s_{e}, e\in E) \in \mathcal{S}$  is a binary vector indicating that the links $e$ with $s_{e}=1$ can be activated simultaneously. The structure of the activation set $\mathcal{S}$ depends on the interference model. 
Under the primary interference constraint, the set $\mathcal{S}$ consists of all binary vectors corresponding to matchings of the underlying graph $\mathcal{G}$~\cite{west2001introduction}, see Fig.~\ref{network}.
 In the case of a wired network, $\mathcal{S}$ is the set of all binary vectors since there is no interference. 
In this paper, we allow an arbitrary link-activation set $\mathcal{S}$, which captures different wireless interference models. 
Let $r\in V$ be the source node at which the broadcast traffic is generated. We consider a time-slotted system. In slot $t$, the number of packets generated at the node $r$ is denoted by $A(t)$, where $A(t)$ is i.i.d. over slots with mean $\lambda$.
These packets need to be delivered to all nodes in the wireless network.

\section{Wireless Broadcast Capacity} \label{capacity_section}

 Intuitively, the network supports a broadcast rate $\lambda$ if there exists a scheduling policy under which all network nodes can receive distinct packets at  rate $\lambda$. The broadcast capacity is the maximally supportable broadcast rate in the network.
Formally, we consider a class $\Pi$ of scheduling policies where each policy $\pi\in\Pi$ is a sequence of actions $\{\pi_t\}_{t\geq 1}$ taken in every slot $t$. Each action $\pi_{t}$ consists of two operations: (i) the scheduler activates a subset of links by choosing a  feasible activation vector  $\boldsymbol s\in\mathcal{S}$; (ii) each node $i$ forwards a subset of packets (possibly empty) to node $j$ over an activated link $(i, j)$, subject to the link capacity constraint. The class $\Pi$ includes policies that use all past and future network information, and may forward any subset of packets over a link.
  
To introduce the notion of broadcast capacity, we define the random variable $R_i^{\pi}(t)$ to be the number of distinct packets received by node $i \in V$ from the beginning of time up to time $t$, under a policy $\pi\in \Pi$. The time average $\lim_{T\to \infty} R^{\pi}_i(T)/T$ is  the rate of distinct packets received at  node $i$.
 
\begin{definition}
A policy $\pi$ is called a 
{``broadcast policy of rate $\lambda$''} 
if 
all nodes receive distinct packets at rate $\lambda$, i.e.,
\begin{equation} \label{bcdef}
\lim_{T\to \infty} \frac{1}{T} R^{\pi}_i(T)= \lambda, \ \text{ for all } i\in V, \ \text{w. p. $1$,}
\end{equation}
where $\lambda$ is the packet arrival rate at the source node $r$.\footnote{We can use the following more rigorous condition in~\eqref{bcdef}:
\begin{equation} \label{eq:703}
\min_{i\in V} \liminf_{T\to\infty} \frac{1}{T} R_{i}^{\pi}(T) = \lambda, \quad \text{w. p. $1$},
\end{equation}
under which all results in this paper still hold.}
\end{definition}

\begin{definition} \label{capacity_def}
The broadcast capacity $\lambda^*$ of a wireless network is  the supremum of all arrival rates $\lambda$ for which there exists a broadcast policy $\pi$ of rate $\lambda$, $\pi\in\Pi$.
\end{definition}

\subsection{An upper bound on broadcast capacity $\lambda^*$} \label{broadcast_ub_section}
We characterize the broadcast capacity $\lambda^*$ of a wireless network by providing a useful upper bound. This upper bound is understood as a necessary cut-set bound of an associated edge-capacitated  graph that  reflects the time-average behavior of the wireless network. We provide an intuitive explanation of the bound that will be formalized in Theorem \ref{broadcast_ub} as follows. 

\begin{figure} [t!] 
\centering
\begin{overpic}[width=0.23\textwidth]{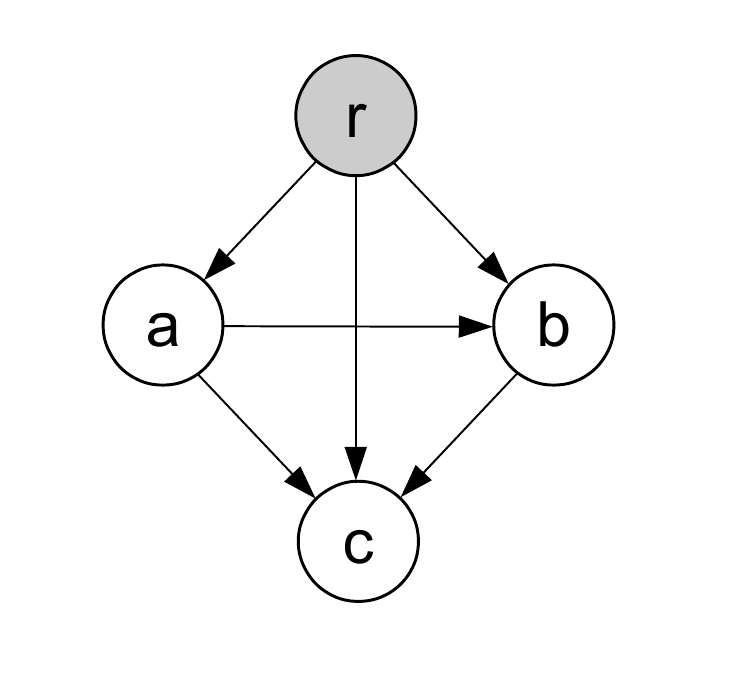}
  \put(25,66){\footnotesize $1/2$}
  \put(63,31){\footnotesize $1/2$}
  \put(52,53){\footnotesize $1/4$}
  \put(23,31){\footnotesize $1/4$}
    \put(62,66){\footnotesize $1/4$}
      \put(36,36){\footnotesize $1/4$}
  \end{overpic}
  \caption{The edge-capacitated graph $\widehat{\mathcal{G}}$ for the wireless network with unit link capacities in Fig.~\ref{network} and under the time-average vector $\boldsymbol\beta^{\pi}=(1/2,1/4,1/4)$. The link weights are the capacities $c_{e} \beta_{e}^{\pi}$. The minimum proper cut in this graph has value $1/2$ (when $U = \{r,a,c\}$ or $\{r,b,c\}$). An upper bound on the  broadcast capacity is obtained by  maximizing this value over all vectors $\boldsymbol\beta^{\pi} \in\conv{\mathcal{S}}$.}
  \label{cut_figure}
\end{figure}
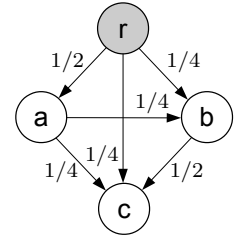

Fix a policy $\pi\in\Pi$. Let $\beta_e^{\pi}$ be the fraction of time link $e\in E$ is activated under $\pi$; that is, we define the vector
\begin{equation} \label{eq:101}
\bm{\beta}^{\pi} = (\beta_e^{\pi}, e\in E) = \lim_{T\to \infty}\frac{1}{T}\sum_{t=1}^{T}\bm{s}^{\pi}(t),
\end{equation}
where $\bm{s}^{\pi}(t)$ is the link-activation vector under policy $\pi$ in slot $t$ (assuming all limits exist). 
The average flow rate over a link $e$ under policy $\pi$ is upper bounded by the product of the link capacity and the fraction of time the link $e$ is activated, i.e., $c_{e} \beta_{e}^{\pi}$.
It is convenient to define an edge-capacitated graph 
$\widehat{\mathcal{G}}=(V, E,(\widehat{c}_{e}))$ 
associated with policy $\pi$, where each directed link $e\in E$ has capacity $\widehat{c}_{e} = c_{e} \beta_{e}^{\pi}$; see Fig.~\ref{cut_figure} for an example of such an edge-capacitated graph.
Next, we provide a bound on the broadcast capacity by  maximizing the broadcast rate at each node on  the graph $\widehat {\cal G}$ over all feasible vectors $\bm{\beta}^{\pi}$.

We define a \emph{proper cut} $U$ of the network graph $\mathcal{G}$ (or $\widehat{\mathcal{G}})$ as a proper subset of the node set $V$ that contains  the source node $r$. Define the link subset
\begin{equation} \label{eq:604}
E_{U} = \{(i, j)\in E \mid i\in U, \ j\notin U\}.
\end{equation}
Since $U \subset V$, there exists a node $n\in V\setminus U$. 
 Consider the throughput of node $n$ under policy $\pi$. The max-flow min-cut theorem shows that the throughput of node $n$  cannot exceed the total link capacity $\sum_{e\in E_{U}} c_{e} \, \beta_{e}^{\pi}$ across the cut $U$. Since the achievable broadcast rate $\lambda^{\pi}$ of policy $\pi$ is a lower bound on the throughput of all nodes, we have $\lambda^{\pi} \leq \sum_{e\in E_{U}} c_{e} \, \beta_{e}^{\pi}$. This inequality holds for all proper cuts $U$ and we have
\begin{equation} \label{eq:102}
\lambda^{\pi} \leq \min_{\text{$U$: a proper cut}}\, \sum_{e\in E_{U}} c_{e} \, \beta_{e}^{\pi}.
\end{equation}
Equation~\eqref{eq:102} holds for any policy $\pi\in\Pi$. Thus, the broadcast capacity $\lambda^{*}$ of the wireless network satisfies
\begin{align*}
\lambda^{*} = \sup_{\pi\in\Pi} \lambda^{\pi} &\leq \sup_{\pi\in\Pi} \min_{\text{$U$: a proper cut}}\, \sum_{e\in E_{U}} c_{e} \, \beta_{e}^{\pi} \\ 
&\leq \max_{\bm{\beta} \in \conv{\mathcal{S}}} \min_{\text{$U$: a proper cut}}\, \sum_{e\in E_{U}} c_{e}\, \beta_{e},
\end{align*}
where the last inequality holds because the vector $\bm{\beta}^{\pi}$ associated with any policy $\pi\in\Pi$ lies in the convex hull $\conv{\mathcal{S}}$ of the activation set $\mathcal{S}$. 
The next theorem formalizes the above characterization of the broadcast capacity $\lambda^*$ of a wireless network.
\begin{theorem} \label{broadcast_ub}
The broadcast capacity $\lambda^{*}$ of a wireless network under general interference constraints satisfies
\begin{equation} \label{eq:602}
\lambda^{*} \leq \max_{\bm{\beta} \in \conv{\mathcal{S}}} \min_{\text{\emph{$U$: a proper cut}}}\, \sum_{e\in E_{U}} c_{e}\, \beta_{e}.
\end{equation}
\end{theorem}
\begin{IEEEproof}[Proof of Theorem~\ref{broadcast_ub}]
See Appendix~\ref{broadcast_ub_proof}.
\end{IEEEproof}


\subsection{In-order packet delivery} \label{constraint1}

Studying the performance of a general broadcast policy is difficult because packets are replicated across the network and may be received out of order. 
To avoid retransmissions of a packet, the nodes must keep track of the identity of the received packets, which complicates the system state---instead of the number of packets received, the system state is described by the subset of packets received at all nodes.

To simplify the system state, we focus on the subset of policies that enforce the \emph{in-order packet delivery} constraint: 
\begin{constraint}[In-order packet delivery]\label{con:1}
A network node is allowed to receive a packet $p$  only if all packets $\{1, 2, \ldots, p-1\}$ have been received by that node. 
\end{constraint}
We denote this policy subclass by $\Pi_{\text{in-order}}$. In-order packet delivery is useful in live media streaming applications~\cite{lstream}, in which buffering out-of-order packets incurs increased delay that degrades video quality. In-order packet delivery greatly simplifies the network state space.   
Let $R_i(t)$ be the number of distinct packets received by node $i$ by time $t$. 
 For policies in $\Pi_{\text{in-order}}$,  the set of received packets by time $t$ at node $i$ is $\{1,\dots, R_i(t)\}$.  
  Therefore, the network state in slot $t$ is completely represented by the vector $\bm{R}(t)=\big(R_i(t), i\in V\big)$.

Imposing the in-order packet delivery constraint potentially reduces the wireless broadcast capacity. It is of interest to study under what conditions in-order packet delivery can attain the broadcast capacity $\lambda^{*}$ of the general policy class $\Pi$.  
The next lemma shows that there exists a cyclic network topology in which any broadcast policy enforcing the in-order packet delivery constraint is not throughput optimal.
\begin{lemma} \label{in_order}
Let ${\lambda^*}_{\text{in-order}}$ be the broadcast capacity of the policy subclass $\Pi_{\text{in-order}} \subset \Pi$ that enforces in-order packet delivery. There exists a network topology with directed cycles such that ${\lambda^*}_{\text{in-order}} < \lambda^{*}$.
\end{lemma}
\begin{IEEEproof}[Proof of Lemma~\ref{in_order}]
See Appendix \ref{in_order_proof}.
\end{IEEEproof}


\subsection{Achieving broadcast capacity of a DAG} \label{sec:601}

In the rest of the paper, we study optimal broadcast algorithms when the network topology is a DAG. For this, we simplify the upper bound~\eqref{eq:602} on the broadcast capacity $\lambda^{*}$ in Theorem \ref{broadcast_ub} in the case of a DAG.
For each receiver node $v\neq r$, we consider the proper cut $U_v$ that separates the network from node $v$:
\begin{equation}\label{eq:pcutsrec}
U_v=V\setminus \{v\}.
\end{equation}
Using these cuts $\{U_{v}, v\neq r\}$, we define another upper bound $\lambda_{\text{DAG}}$ on the broadcast capacity $\lambda^{*}$ as:
\begin{align}  \label{eq:DAGcap}
\lambda_{\text{DAG}}&\triangleq \max_{\bm{\beta}\in \conv{\mathcal{S}}}\min_{\{U_v, v\neq r\} }\, \sum_{e\in E_{U_{v}}} c_{e}\, \beta_{e}\\
&\geq \max_{\bm{\beta} \in \conv{\mathcal{S}}} \min_{\text{$U$: a proper cut}}\, \sum_{e\in E_{U}} c_{e}\, \beta_{e}\geq \lambda^{*}, \notag
\end{align}
where the first inequality uses the subset relation $\{U_v, v\neq r\}\subseteq \{\text{$U$: a proper cut}\}$ and the second inequality follows from Theorem \ref{broadcast_ub}. In Section \ref{sec:algorithm}, we will propose a dynamic policy that belongs to the policy class $\Pi_{\text{in-order}}$ and achieves any arrival rate that is less than $\lambda_{\text{DAG}}$. Combining this result with~\eqref{eq:DAGcap}, we establish that the broadcast capacity of a DAG is
 \begin{align} 
 \lambda^*=\lambda_{\text{DAG}} &=\max_{\bm{\beta}\in \conv{\mathcal{S}}}\min_{\{U_v, v\neq r\} }\, \sum_{e\in E_{U_{v}}} c_{e}\, \beta_{e}, \notag \\
 &= \max_{\bm{\beta} \in \conv{\mathcal{S}}} \min_{\text{$U$: a proper cut}}\, \sum_{e\in E_{U}} c_{e}\, \beta_{e}, \label{eq:603}
 \end{align}
which is achieved by a broadcast policy that uses in-order packet delivery. In other words, imposing the in-order packet delivery constraint does not reduce the broadcast capacity when the network topology is a DAG.

\section{DAG Broadcast Algorithm}\label{sec:algorithm}

We design an optimal broadcast policy for a wireless DAG. We start with imposing an additional constraint that leads to a new subclass of policies in which it is possible to describe the network dynamics by means of relative packet deficits, i.e., $R_i(t)-R_j(t)$, between nodes $i$ and $j$.
We analyze the dynamics of the minimum relative packet deficit at each node~$j$, where the minimization is over all incoming neighbors of $j$. The minimum relative deficits play the role of virtual queues in the system, and we design a control policy that stabilizes them. The main result of this section is to show that this control policy achieves the broadcast capacity whenever the network topology is a DAG.

\subsection{System state by means of packet deficits}

We show in Section~\ref{constraint1} that, in the policy class $\Pi_{\text{in-order}}$, the system state is represented by the vector $\bm{R}(t)$.
To simplify the system dynamics further, we impose another constraint on $\Pi_{\text{in-order}}$ as follows. We say that node $i$ is an \emph{in-neighbor} of node $j$ if there exists a directed link $(i, j)\in E$ in the underlying graph $\mathcal{G}$.
\begin{constraint}\label{con:2}
A packet $p$ is eligible for  transmission to node $j$ in a slot only if all the in-neighbors of  $j$ have received packet $p$ in previous slots.
 \end{constraint}

We denote this new policy class by $\Pi^{*} \subseteq \Pi_{\text{in-order}}$.\footnote{If the network contains a directed cycle, then a deadlock may occur under a policy in $\Pi^{*}$ and yields zero broadcast throughput. This problem does not arise when the network is a DAG.}
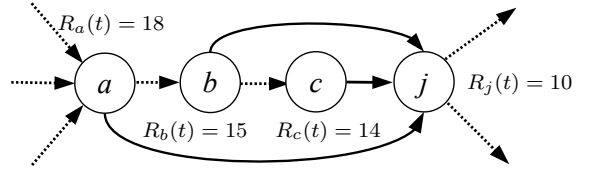
\begin{figure} [h!] 
\centering
\begin{overpic}[width=0.39\textwidth]{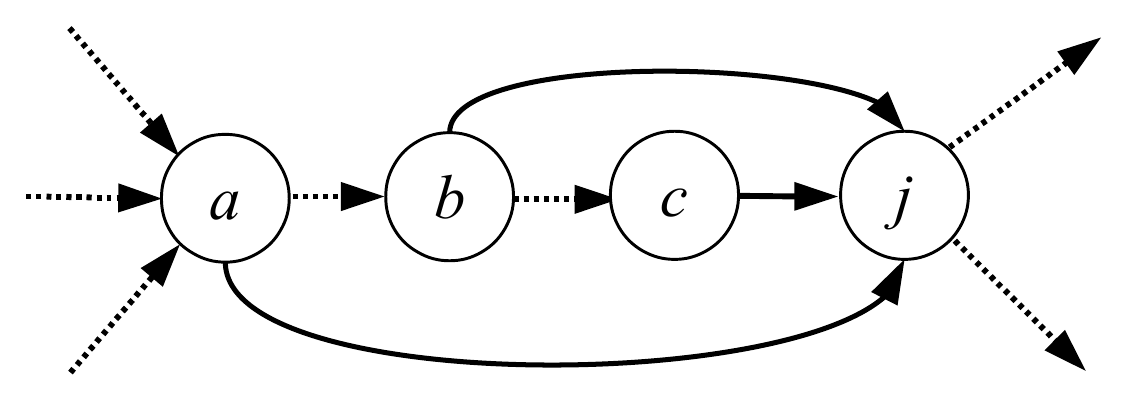}
  \put(11,28){\footnotesize $R_a(t)=18$}
  \put(27,8){\footnotesize $R_b(t)=15$}
  \put(52,8){\footnotesize $R_c(t)=14$}
  \put(88,17){\footnotesize $R_j(t)=10$}
  \end{overpic}
  \caption{Under a policy $\pi\in \Pi^*$, the set of packets available for transmission to node $j$ in slot $t$ is $\{11,12,13,14\}$,
  which are available at all in-neighbors of node $j$.
  The in-neighbor of $j$ inducing the smallest packet deficit is $i^*_t=c$, and $X_{j}(t) = \min\{Q_{aj}(t), Q_{bj}(t), Q_{cj}(t)\} = 4$. }
  \label{pi*_figure}
\end{figure}
 The following properties of the system states $\bm{R}(t)$ under a policy in $\Pi^{*}$ are useful.
\begin{lemma} \label{Q_positivity_lemma}
Let $\text{In}(j)$ denote the set of in-neighbors of a node $j$ in the network. 
Under any policy $\pi\in\Pi^{*}$, we have:
\begin{enumerate}
\item[(1)] $R_{j}(t) \leq \min_{i\in \text{In}(j)} R_{i}(t)$ for all nodes $j\neq r$.
\item[(2)] The collection of packets that are eligible to be received by a node $j \neq r$ from its in-neighbors in slot $t$ is
\[
\big\{p \mid R_{j}(t)+1 \leq p \leq \min_{i\in\text{In}(j)} R_{i}(t)\big\}.
\]
\end{enumerate}
\end{lemma}

We define the \emph{packet deficit} over a directed link $(i,j)\in E$ by $Q_{ij}(t) = R_i(t)-R_j(t)$. Under a policy in $\Pi^*$, $Q_{ij}(t)$ is always nonnegative because, by part (1) of Lemma~\ref{Q_positivity_lemma}, we get
\[
Q_{ij}(t) = R_{i}(t) - R_{j}(t) \geq \min_{k\in\text{In}(j)} R_{k}(t) - R_{j}(t) \geq 0.
\]
The quantity $Q_{ij}(t)$ is the number of packets that have been received by a node $i$ but not by node~$j$.
Intuitively, if all packet deficits $Q_{ij}(t)$ are bounded, then the total number of packets received by any node is not far from that generated at  the source node $r$; as a result, the broadcast throughput is equal to the packet arrival rate.

To analyze 
 the system dynamics under a policy in $\Pi^{*}$, it is useful to define the \emph{minimum packet deficit} at node $j$ by
\begin{equation} \label{X_def}
X_j(t) = \min_{i\in \text{In}(j)} Q_{ij}(t), \quad \forall j\neq r.
\end{equation}
From part~(2) of Lemma~\ref{Q_positivity_lemma}, $X_j(t)$ is the maximum number of packets that node $j$ is allowed to receive from its in-neighbors in slot $t$. As an example, Fig.~\ref{pi*_figure} shows that the packet deficits at node $j$, as compared to the upstream nodes $a$, $b$, and $c$, are $Q_{aj}(t)=8$, $Q_{bj}(t)=5$, and $Q_{cj}(t)=4$, respectively. Thus $X_{j}(t)=4$ and node $j$ is only allowed to receive four packets in slot $t$ due to Constraint \ref{con:2}.
We can rewrite $X_{j}(t)$ as
\begin{equation} \label{eq:115}
X_{j}(t) = Q_{i_{t}^{*}j}(t), \quad \text{where } i_{t}^{*} = \arg\min_{i \in \text{In}(j)} Q_{ij}(t),
\end{equation}
and the node $i_{t}^{*}$ is the in-neighbor of node $j$ from which node $j$ has the smallest packet deficit in slot $t$; ties are broken arbitrarily in deciding $i_{t}^{*}$.\footnote{We note that the minimizer $i_{t}^{*}$ is a function of the node $j$ and the time slot $t$; we slightly abuse the notation by neglecting $j$ to avoid clutter.} Our optimal broadcast policy will use the minimum packet deficits $X_{j}(t)$.


\subsection{The dynamics of the system variable $X_{j}(t)$}
We analyze the dynamics of the system variables
\begin{equation} \label{eq:104}
X_j(t) = Q_{i_{t}^{*}j}(t) = R_{i_{t}^{*}}(t)-R_{j}(t)
\end{equation}
under a policy $\pi \in \Pi^*$. 
Define the service rate vector $\bm{\mu}(t) = (\mu_{ij}(t))_{i, j\in V}$ by
\[
\mu_{ij}(t) = \begin{cases} c_{ij} & \text{if $(i, j)\in E$ and the link (i, j) is activated,} \\ 0 & \text{otherwise.} \end{cases}
\]
Equivalently, we may write $\mu_{ij}(t) = c_{ij} s_{ij}(t)$, and the number of packets forwarded over a link is constrained by the choice of the link-activation vector $\bm{s}(t)$. At node $j$,  the increase in the value of $R_j(t)$ depends on the identity of the received packets; in particular, node $j$ must receive distinct packets. Next, we clarify which packets are to be received by node $j$.
 
The number of available packets for reception at node $j$ is $\min\{X_{j}(t), \sum_{k\in V} \mu_{kj}(t)\}$, because: (i) $X_{j}(t)$ is the maximum number of packets node $j$ can receive from its in-neighbors due to Constraint \ref{con:2}; (ii) $\sum_{k\in V} \mu_{kj}(t)$ is the total incoming transmission rate at node $j$ under a given link-activation decision. 
To correctly derive the dynamics of $R_j(t)$, we consider the following efficiency requirement on policies in $\Pi^*$:
\begin{constraint}[Efficient forwarding]\label{con:3}
Given a service rate vector $\bm{\mu}(t)$, node $j$ pulls from the activated incoming links the following subset of packets 
\begin{equation} \label{eq:701}
\Big\{p \mid R_j(t)+1\leq p\leq R_j(t)+\min\{X_{j}(t), \sum_{k\in V} \mu_{kj}(t)\}\Big\},
\end{equation}
where which packets are pulled over each incoming link are arbitrary but must satisfy Constraint~\ref{con:1}.\footnote{Due to Constraints \ref{con:1} and~\ref{con:2}, the packets in~\eqref{eq:701} have been received by all in-neighbors of node $j$.}
\end{constraint}
Constraint~\ref{con:3} states that scheduling policies must avoid forwarding the same packet to a node over two different incoming links in a slot. Under certain interference models such as the primary interference model, at most one incoming link is activated at a node in a slot, and Constraint \ref{con:3} is redundant.

In~\eqref{eq:104}, the packet deficit $Q_{i_{t}^{*}j}(t)$ increases with $R_{i_{t}^{*}}(t)$ and decreases as $R_{j}(t)$ increases, where $R_{i_{t}^{*}}(t)$ and $R_{j}(t)$ are both non-decreasing. It follows that we can upper bound the increase of $Q_{i_{t}^{*}j}(t)$ 
 by the total capacity $\sum_{m\in V} \mu_{mi_{t}^{*}}(t)$ of activated incoming links at node~$i_{t}^{*}$. 
 Also, we can express the decrease of $Q_{i_{t}^{*}j}(t)$  by the exact number of distinct packets received by node $j$ from its in-neighbors, and it is given by $\min\{X_{j}(t), \sum_{k\in V} \mu_{kj}(t)\}$ by Constraint \ref{con:3}. Consequently, the one-slot evolution of the variable $Q_{i_{t}^{*}j}(t)$ is given by\footnote{We emphasize that the node $i_{t}^{*}$ is defined in~\eqref{eq:115}, depends on the particular node $j$ and time $t$, and may be different from the node $i_{t+1}^{*}$.}
\begin{align}
Q_{i_t^*j}(t+1) &\ \leq \big(Q_{i_t^*j}(t) - \sum_{k\in V} \mu_{kj}(t)\big)^+  + \sum_{m\in V}\mu_{m i_t^*}(t) \notag \\
&\ \leq \big(X_j(t)  - \sum_{k\in V} \mu_{kj}(t)\big)^+ + \sum_{m\in V}\mu_{mi_t^*}(t), \label{eq:105}
\end{align}
where $(x)^{+} = \max(x, 0)$ and we recall that $X_j(t)=Q_{i_t^*j}(t)$. It follows that $X_{j}(t)$ evolves over slot $t$ according to
\begin{align}\label{bnd2}
 X_j(t+1) &\stackrel{(a)}{=} \min_{i\in \text{In}(j)} Q_{ij}(t+1) \stackrel{(b)}{\leq} Q_{i_t^*j}(t+1) \notag\\
&\stackrel{(c)}{\leq} \big(X_j(t)  - \sum_{k\in V} \mu_{kj}(t)\big)^+ + \sum_{m\in V} \mu_{mi_t^*}(t),
\end{align}
where (a) follows the definition of $X_{j}(t)$, (b) follows because node $i_{t}^{*} \in \text{In}(j)$, and (c) follows from~\eqref{eq:105}. In~\eqref{bnd2}, we abuse the notation to define $\sum_{m\in V} \mu_{mr}(t) = A(t)$ for the source node $r$, where $A(t)$ is the number of exogenous packet arrivals in slot $t$.

\subsection{The optimal broadcast policy} \label{lyapunov}

Our broadcast policy is designed to keep the minimum deficits $X_{j}(t)$ bounded.
For this, we regard the variables $X_{j}(t)$ as virtual queues that follow the dynamics~\eqref{bnd2}. By performing drift analysis on the virtual queues $X_{j}(t)$, we propose the following max-weight-type broadcast policy that belongs to the policy subclass $\Pi^{*}$ which enforces Constraints \ref{con:1}, \ref{con:2}, and \ref{con:3}.
 We will show that this policy achieves the broadcast capacity $\lambda^{*}$ of a wireless network over the general policy class $\Pi$ when the underlying network graph is a DAG.

\noindent \rule[0.05in]{3.5in}{0.01in}

\textbf{Optimal Broadcast Policy $\pi^{*}$ over a Wireless DAG:}
\\ \phantom{a}

In slot $t$, the algorithm has the input $\{R_{j}(t), j\in V\}$ and performs the following four steps.

\textbf{Step 1:}
 For each link $(i, j)\in E$, compute the deficit $Q_{ij}(t) = R_{i}(t) - R_{j}(t)$ and the set of nodes $K_{j}(t)$ for which node $j$ is the deficit minimizer:
\begin{equation} \label{eq:110}
K_{j}(t) = \big\{k\in V\mid j = \arg\min_{m\in\text{In}(k)} Q_{mk}(t)\big\}.
\end{equation}
We note that the ties are broken arbitrarily in finding a deficit minimizer.

\textbf{Step 2:} Compute $X_{j}(t) = \min_{i\in\text{In}(j)} Q_{ij}(t)$ for $j\neq r$ and assign to link $(i, j)$ the weight
\begin{equation} \label{eq:111}
W_{ij}(t) = \big(X_{j}(t) - \sum_{k\in K_{j}(t)} X_{k}(t)\big)^{+},
\end{equation}
where $W_{ij}(t)$ is the minimum deficit of node $j$ minus that of all nodes for which node $j$ is the deficit minimizer. Intuitively, the term $W_{ij}(t)$ arises because while delivering a packet to node $j$ decreases $X_{j}(t)$ by one, it also increases $X_{k}(t)$ by one for all nodes for which node $j$ is an in-neighbor and the deficit minimizer.

\textbf{Step 3:} In slot $t$, choose the link-activation vector $\bm{s}(t) = (s_{e}(t), e\in E)$ such that
\begin{equation} \label{eq:601}
\bm{s}(t)  \in \arg\max_{ (s_{e}, e\in E)\in \mathcal{S}} \sum_{e\in E} c_{e} s_{e} W_{e}(t).
\end{equation}
Every node $j\neq r$ uses activated incoming links to pull  packets $\{R_j(t)+1, \dots,R_j(t)+ \min\{\sum_ic_{ij} s_{ij}(t), X_{j}(t)\}\}$ from its in-neighbors according to Constraint \ref{con:3}.

\textbf{Step 4:} The vector $(R_{j}(t), j\in V)$ is updated as follows:
\[
R_{j}(t+1) = \begin{cases} R_{j}(t) + A(t), & j = r, \\ R_{j}(t) + \min\{\sum_ic_{ij} s_{ij}(t), X_{j}(t)\}, & j\neq r, \end{cases}
\]
and $R_{j}(0)=0$ for all $j\in V$.

\noindent \rule[0.05in]{3.5in}{0.01in}

\begin{figure}[t!] 
\subfigure{
\label{fig:601}
\begin{overpic}[width=0.23\textwidth]{Network_2b}
\put(41,92){\scriptsize \textbf{Step 1}}
\put(57,80){\scriptsize $R_r(t)=10$}
\put(-1,61){\scriptsize $R_a(t)=3$}
\put(67,61){\scriptsize $R_b(t)=3$}
\put(58,21){\scriptsize $R_c(t)=2$}
\put(102,81){\scriptsize $Q_{ra}(t)=7$}
\put(102,71){\scriptsize $Q_{rb}(t)=7$}
\put(102,61){\scriptsize $Q_{rc}(t)=8$}
\put(102,51){\scriptsize $Q_{ab}(t)=0$}
\put(102,41){\scriptsize $Q_{ac}(t)=1$}
\put(102,31){\scriptsize $Q_{bc}(t)=1$}
\put(142,81){\scriptsize $K_{r}(t)=\{a\}$}
\put(142,71){\scriptsize $K_{a}(t)=\{b,c\}^*$}
\put(142,61){\scriptsize $K_{b}(t)=\{\emptyset\}$}
\put(142,51){\scriptsize $K_{c}(t)=\{\emptyset\}$}
\end{overpic}
}
\subfigure{
\label{fig:602}
\begin{overpic}[width=0.23\textwidth]{Network_2b}
\put(41,92){\scriptsize \textbf{Step 2}}
\put(-1,61){\scriptsize $X_a(t)=7$}
\put(85,71){\scriptsize $X_b(t)=0$}
\put(58,21){\scriptsize $X_c(t)=1$}
\put(70,81){\scriptsize $W_{ra}(t)=(X_a(t)-X_b(t)-X_c(t))^+=6$}
\put(120,71){\scriptsize $W_{rb}(t)=(X_b(t))^+=0$}
\put(120,61){\scriptsize $W_{rc}(t)=(X_c(t))^+=1$}
\put(120,51){\scriptsize $W_{ab}(t)=(X_b(t))^+=0$}
\put(120,41){\scriptsize $W_{ac}(t)=(X_c(t))^+=1$}
\put(120,31){\scriptsize $W_{bc}(t)=(X_c(t))^+=1$}
\end{overpic}
}
\subfigure{
\label{fig:603}
\begin{overpic}[width=0.23\textwidth]{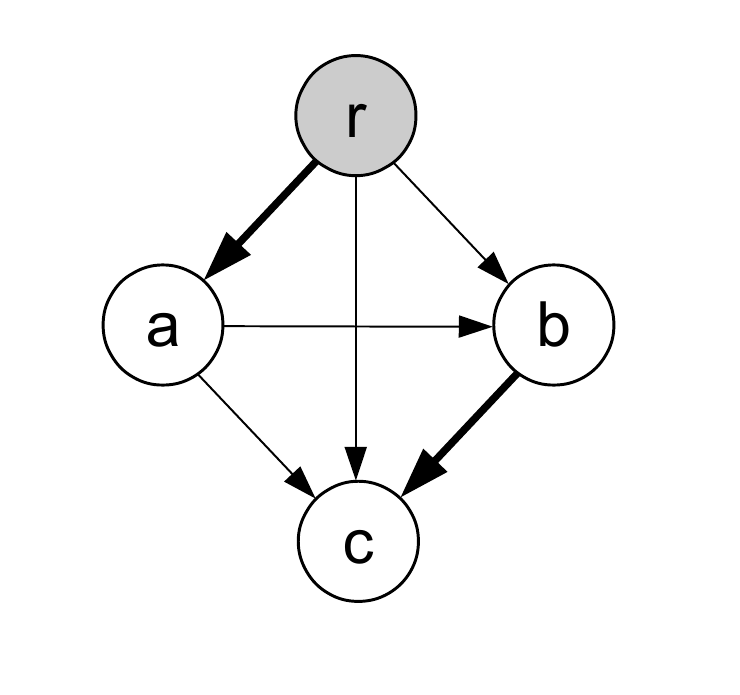}
\put(41,92){\scriptsize \textbf{Step 3}}
\put(-1,36){\scriptsize $R_a(t)=3$}
\put(58,16){\scriptsize $R_c(t)=2$}
\put(102,81){\scriptsize $\mathbf{s}_1$: $W_{ra}(t)+W_{bc}(t)=7$}
\put(102,71){\scriptsize $\mathbf{s}_2$: $W_{rb}(t)+W_{ac}(t)=1$}
\put(102,61){\scriptsize $\mathbf{s}_3$: $W_{rc}(t)+W_{ab}(t)=1$}
\put(102,51){\scriptsize Choose the link-activation vector $\mathbf{s}_1$}
\put(102,41){\scriptsize Forward the next packet \#4 on $(r,a)$}
\put(102,31){\scriptsize Forward the next packet \#3 on $(b,c)$}
\put(29,67){\footnotesize  \#4}
\put(63,29){\footnotesize  \#3}
\end{overpic}
}
\subfigure{
\label{fig:604}
\begin{overpic}[width=0.23\textwidth]{Network_2b}
\put(41,92){\scriptsize \textbf{Step 4}}
\put(57,80){\scriptsize $R_r(t+1)=11$}
\put(-6,65){\scriptsize $R_a(t+1)=4$}
\put(65,62){\scriptsize $R_b(t+1)=3$}
\put(58,21){\scriptsize $R_c(t+1)=3$}
\put(112,81){\scriptsize One packet arrives at the source}
 \end{overpic}
}
 \caption{Running the optimal broadcast policy $\pi^{*}$ in slot $t$ in a wireless network with unit-capacity links and under the primary interference constraint. Step 1: computing the deficits $Q_{ij}(t)$ and $K_{j}(t)$; a tie is broken in choosing node $a$ as the in-neighbor deficit minimizer for node $c$, hence $c\in K_{a}(t)$; node $b$ is also a deficit minimizer for node $c$. Step 2: computing $X_{j}(t)$ for $j\neq r$ and $W_{ij}(t)$. Step 3: finding the link activation vector that is a maximizer in~\eqref{eq:601}, and forwarding the next in-order packets over activated links. Step 4: a new packet arrives at the source node $r$, and new values of $\{R_{r}(t+1), R_{a}(t+1), R_{b}(t+1), R_{c}(t+1)\}$ are updated.
 }
 \label{algorithm_fig}
\end{figure}
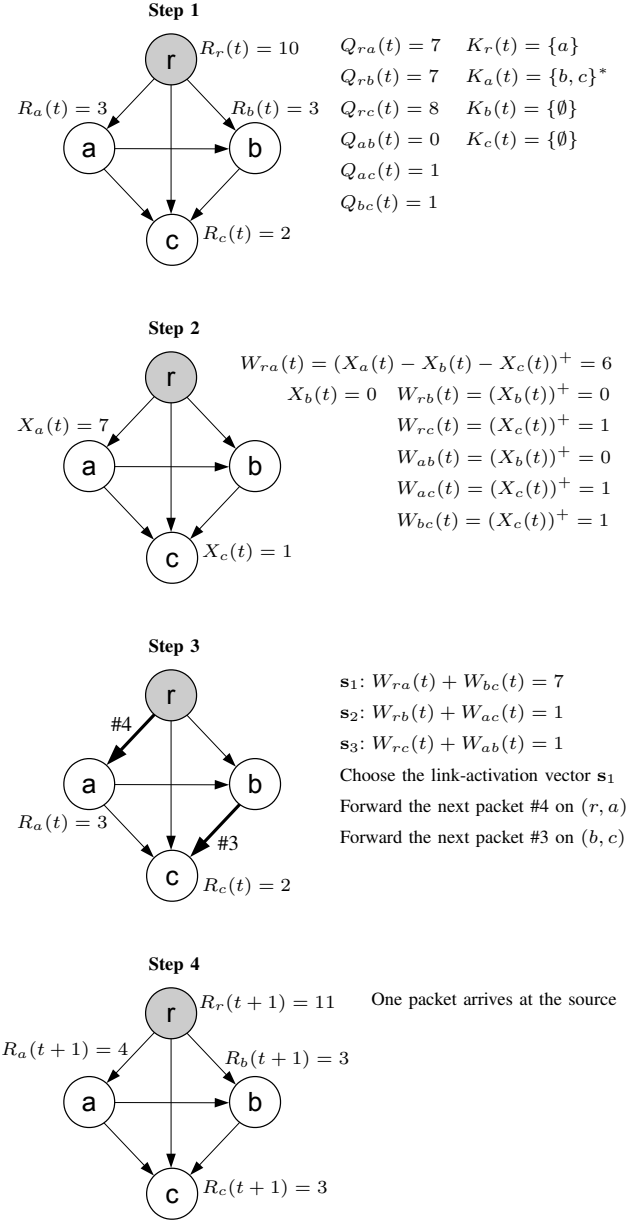

We illustrate the above algorithm in an example in Fig.~\ref{algorithm_fig}. The next theorem demonstrates the optimality of the broadcast policy $\pi^{*}$.
\begin{theorem} \label{main_theorem}
If the underlying network graph $\mathcal{G}$ is a DAG, then for any exogenous packet arrival rate $\lambda <\lambda_{\text{DAG}}$, the broadcast policy $\pi^{*}$ yields
\[
\lim_{T \to \infty}\frac{R_i^{\pi^{*}}(T)}{T} = \lambda, \ \forall i\in V, \quad \text{with probability $1$,}
\]
where $\lambda_{\text{DAG}}$ is the upper bound on the broadcast capacity $\lambda^{*}$ in the general policy class $\Pi$, as shown in~\eqref{eq:DAGcap}. Consequently, the broadcast policy $\pi^{*}$ achieves the broadcast capacity $\lambda^{*}$ when the network topology is a DAG.
\end{theorem}
\begin{IEEEproof}[Proof of Theorem~\ref{main_theorem}]
See Appendix \ref{main_theorem_proof}.
\end{IEEEproof}
Fig.~\ref{policy_fig} shows the relationship between different policy classes discussed in this paper.
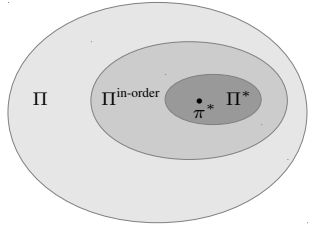
\begin{figure} [htbp] 
\centering
\hspace*{1.5in}
\begin{overpic}[width=0.24\textwidth]{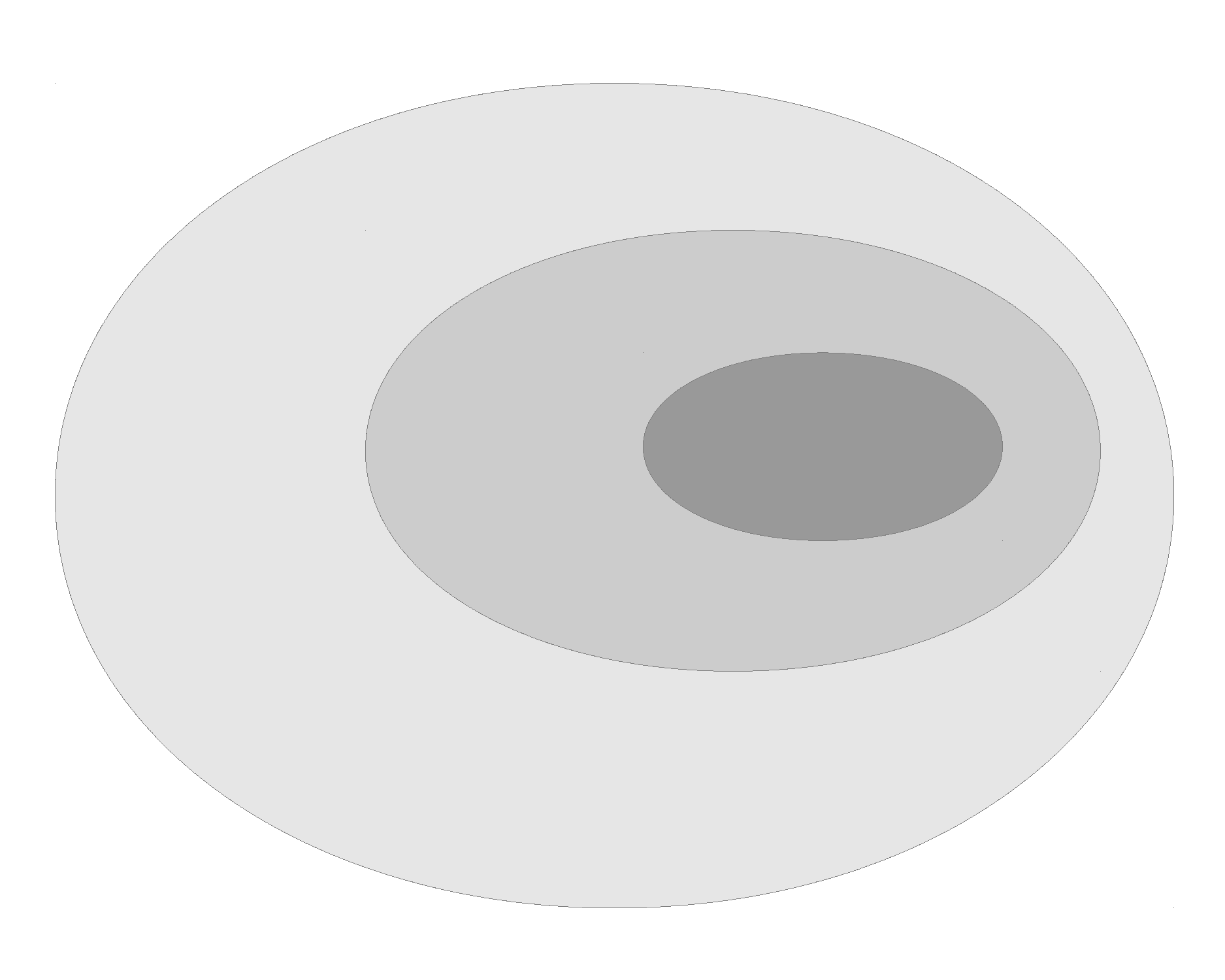}
\put(12,42){\scriptsize $\Pi$}
\put(33,42){\scriptsize  $\Pi^{\text{in-order}}$}
\put(71,42){\scriptsize  $\Pi^*$}
\put(61,38){\scriptsize  $\pi^*$}
\put(63,43){\circle*{2}}
\put(-90,64){\scriptsize  $\Pi$: all policies that perform}
\put(-81,57){\scriptsize   link activations and routing}
\put(-90,44){\scriptsize  $\Pi^{\text{in-order}}$: policies that enforce}
\put(-68,37){\scriptsize in-order packet delivery}
\put(-90,24){\scriptsize  $\Pi^*$: policies that allow reception}
\put(-78,17){\scriptsize only if all in-neighbors have}
\put(-78,10){\scriptsize received the specific packet}
 \end{overpic}
 \caption{The relationship between different policy classes.}
 \label{policy_fig}
\end{figure}

\subsection{Number of disjoint spanning trees in a DAG} \label{sec:disjoint}
Theorem~\ref{main_theorem} provides an interesting combinatorial result that relates the number of disjoint spanning trees in a DAG to the in-degrees of its nodes.
\begin{lemma} \label{lem:701}
Consider a directed acyclic graph $G=(V, E)$ that is rooted at a node $r$, has unit-capacity links, and possibly contains parallel edges. The maximum number $k^{*}$ of disjoint spanning trees in $G$ is given by
\[
k^*=\min_{v\in V\setminus \{r\}}d_{\text{in}}(v),
\]
where $d_{\text{in}}(v)$ denotes the in-degree of the node $v$.
\end{lemma}
\begin{IEEEproof}[Proof of Lemma~\ref{lem:701}]
See Appendix~\ref{pf:701}.
\end{IEEEproof}

\section{Simulation Results}\label{sec:simulations}

We simulate the optimal broadcast policy $\pi^{*}$ in a wireless DAG network with the primary interference constraint in Fig.~\ref{fig:602}; the link capacities are presented as weights on the links. The broadcast capacity $\lambda^{*}$ of the network is upper bounded by the maximum throughput of node $c$, which is $1$ because at most one of its incoming links can be activated at any time. To show that the broadcast capacity is indeed $\lambda^{*} = 1$, we consider the three spanning trees $\{\mathcal{T}_1, \mathcal{T}_2, \mathcal{T}_3\}$ rooted at the source node $r$. By finding the optimal time-sharing of all feasible link activations over a subset of spanning trees using linear programming, we can show that the maximum broadcast throughput using only the spanning tree $\mathcal{T}_{1}$ is $3/4$. The maximum broadcast throughput over the two trees $\{\mathcal{T}_{1}, \mathcal{T}_{2}\}$ is $6/7$, and that over all three trees $\{\mathcal{T}_1, \mathcal{T}_2, \mathcal{T}_3\}$ is $1$. Thus, the upper bound is achieved and the broadcast capacity is $\lambda^{*}=1$.

We compare our broadcast policy $\pi^{*}$ with the tree-based policy $\pi_{\text{tree}}$ in~\cite{swati}. While the policy $\pi_{\text{tree}}$ is originally proposed to transmit multicast traffic in a wired network by balancing traffic over multiple trees, we slightly modify the policy $\pi_{\text{tree}}$ for broadcasting packets over spanning trees in the wireless setting; link activations are chosen according to the max-weight procedure. See Fig.~\ref{fig:601} for a comparison of the average delay performance under the policy $\pi^{*}$ and the tree-based policy $\pi_{\text{tree}}$ over different subset of trees. The simulation duration is $10^{5}$ slots. We observe that the policy $\pi^{*}$ achieves the broadcast capacity $\lambda^{*}=1$ and is throughput optimal.

\begin{figure}[ht!]
\centering
\subfigure[The wireless network]{
\includegraphics[scale=0.4]{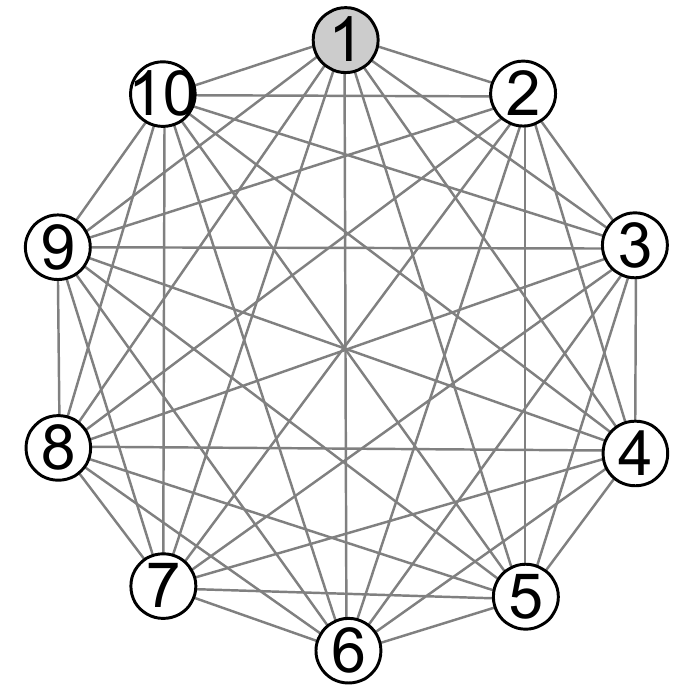}
}
\subfigure[Tree $\mathcal{T}_{1}$]{
\includegraphics[scale=0.4]{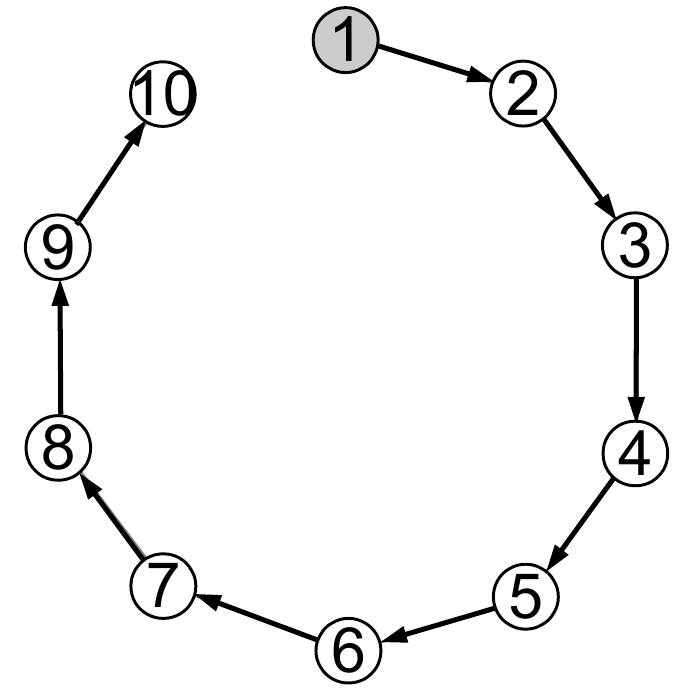}
}
\subfigure[Tree $\mathcal{T}_{2}$]{
\includegraphics[scale=0.4]{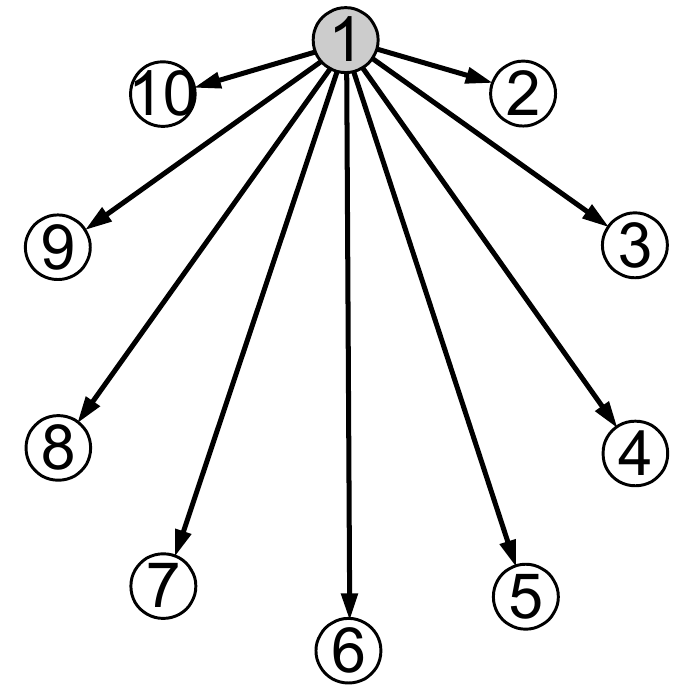}
}
\subfigure[Tree $\mathcal{T}_{3}$]{
\includegraphics[scale=0.4]{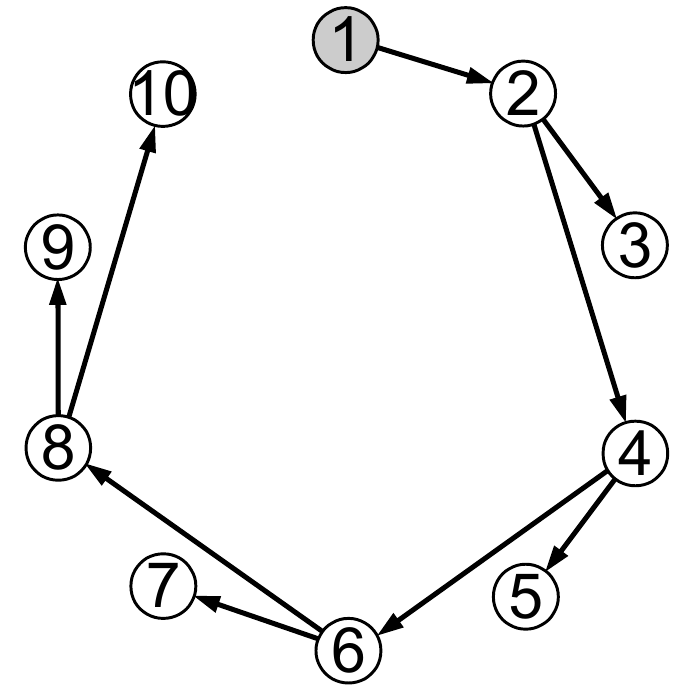}
}
\subfigure[Tree $\mathcal{T}_{4}$]{
\includegraphics[scale=0.4]{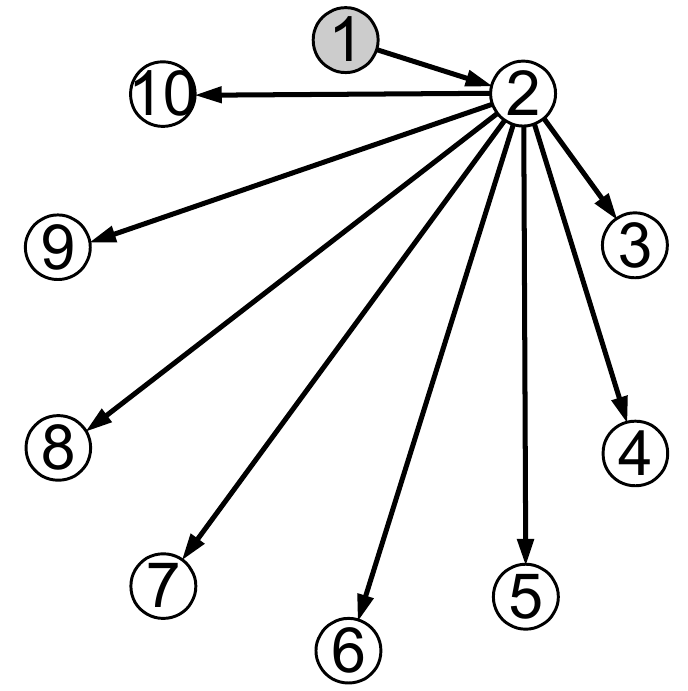}
}
\subfigure[Tree $\mathcal{T}_{5}$]{
\includegraphics[scale=0.4]{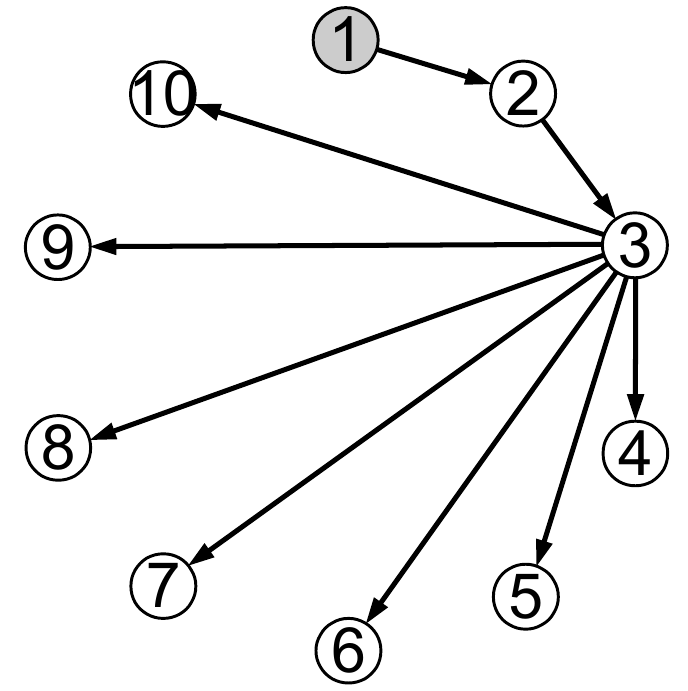}
}
\caption{The $10$-node wireless DAG network and a subset of spanning trees.}
\label{fig:605}
\end{figure}

The broadcast policy $\pi^{*}$ does not rely on the limited tree structures and therefore has the potential to exploit all degrees of freedom in packet forwarding in the network; such freedom may lead to better delay performance as compared to the tree-based policy. To observe this effect, we consider the $10$-node DAG network subject to the primary interference constraint in Fig.~\ref{fig:605}. For every pair of node $\{i,j\}$, $1\leq i <j \leq 10$, the network has a directed link from $i$ to $j$ with capacity $(10-i)$. By induction, we can calculate the number of spanning trees rooted at the source node $1$ to be $9!\approx 3.6\times 10^5$. We choose five arbitrary spanning trees $\{\mathcal{T}_i, 1\leq i \leq 5\}$, over which the tree-based algorithm $\pi_{\text{tree}}$ is simulated. Table~\ref{delay_table} demonstrates the superior delay performance of the broadcast policy $\pi^{*}$, as compared to that of the tree-based algorithm $\pi_{\text{tree}}$ over different subsets of the spanning trees. It also shows that a tree-based algorithm that does not use enough trees would result in degraded throughput.

\begin{figure}
\centering
\subfigure[The wireless network]{
\begin{overpic}[width=0.2\textwidth]{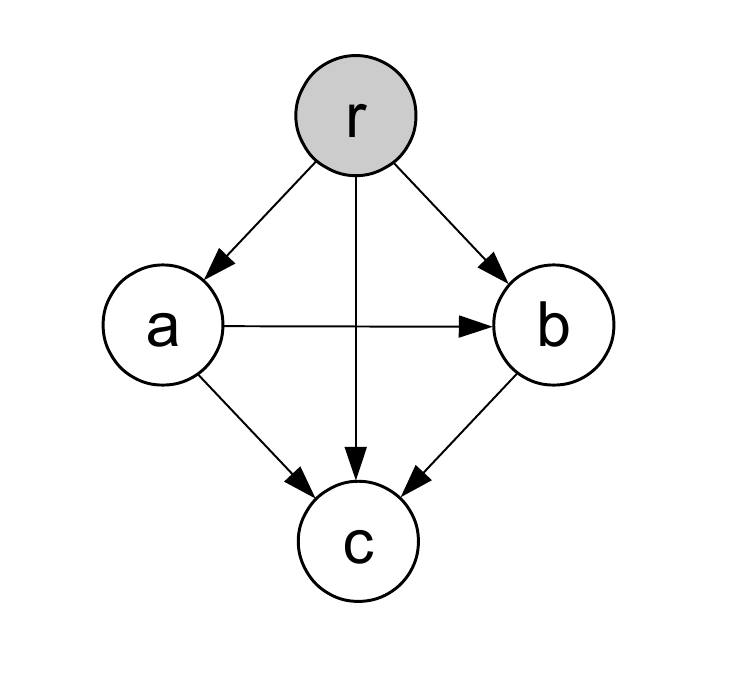}
\put(31,66){\footnotesize $3$}
\put(61,66){\footnotesize $1$}
\put(54,52){\footnotesize $2$}
\put(43,39){\footnotesize $1$}
\put(29,31){\footnotesize $1$}
\put(64,31){\footnotesize $1$}
\end{overpic}
}
\subfigure[Tree $\mathcal{T}_{1}$]{
\includegraphics[width=0.2\textwidth]{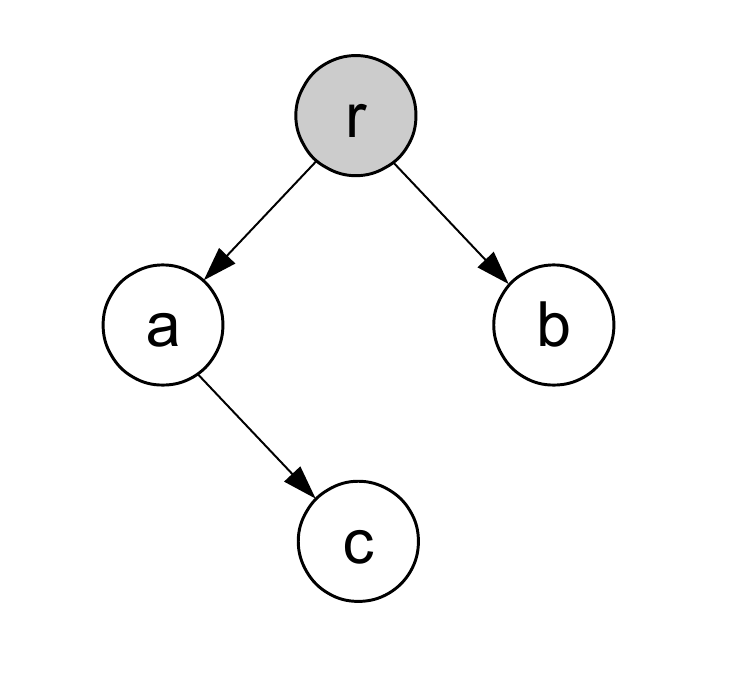}
}
\subfigure[Tree $\mathcal{T}_{2}$]{
\includegraphics[width=0.2\textwidth]{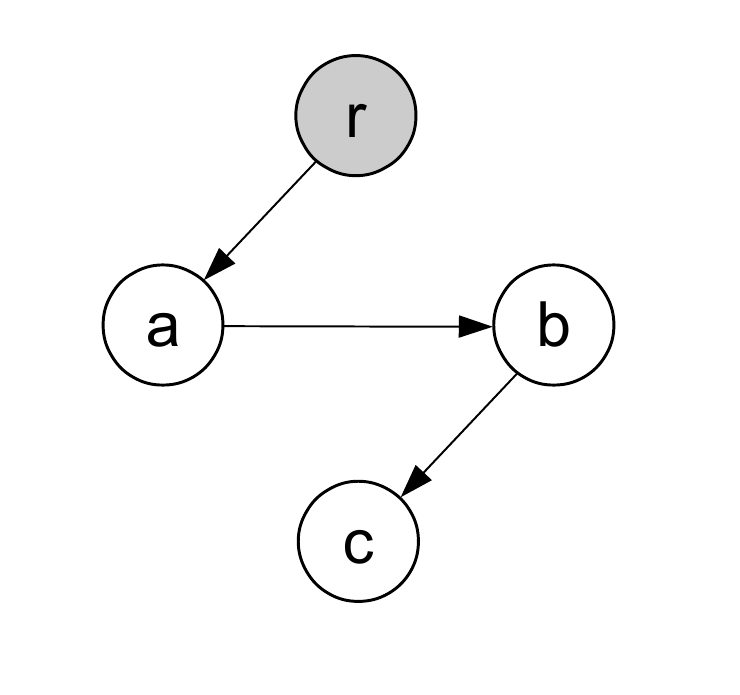}
}
\subfigure[Tree $\mathcal{T}_{3}$]{
\includegraphics[width=0.2\textwidth]{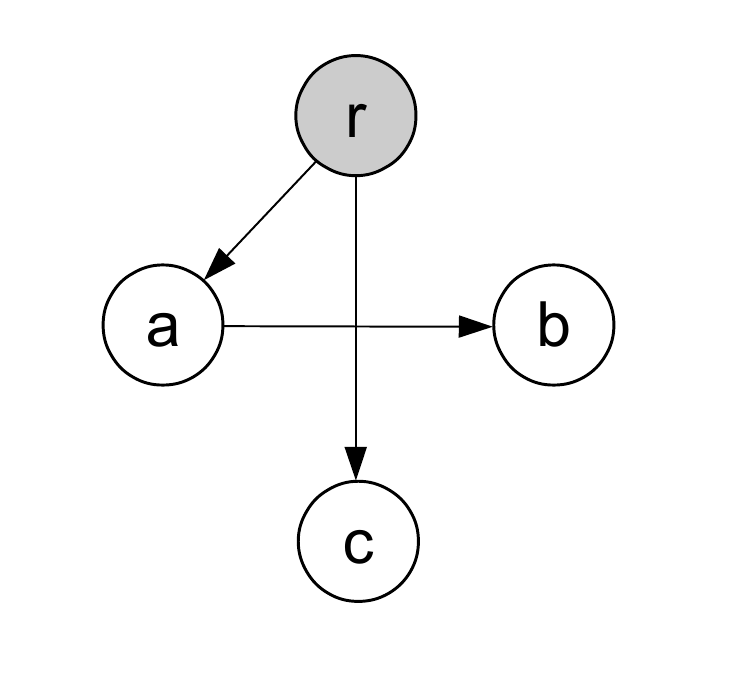}
}
\caption{A wireless DAG network and its three embedded spanning trees.}
\label{fig:602}
\end{figure}

\begin{figure}
\centering
\begin{overpic}[width=0.5\textwidth]{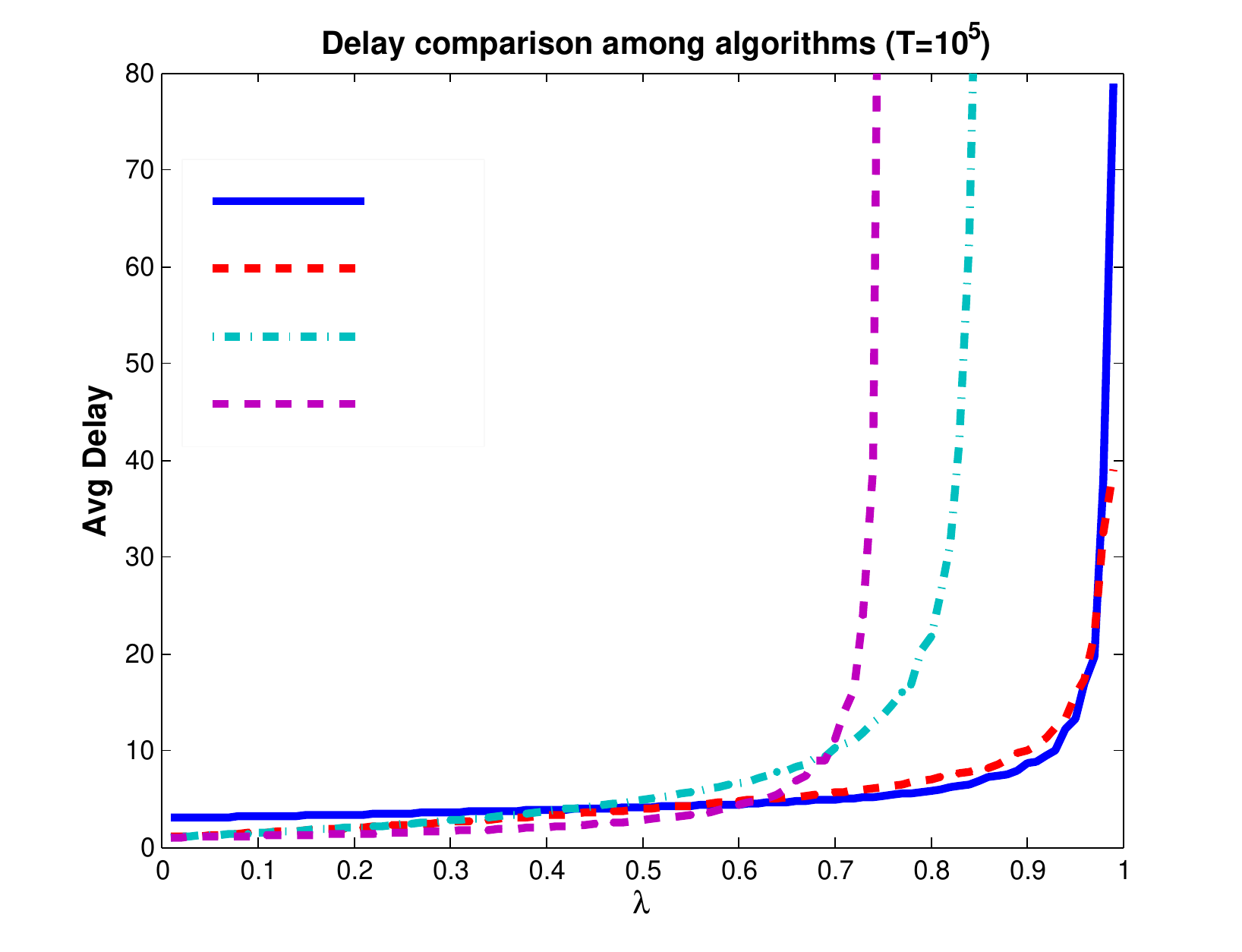}
\put(30,60){\footnotesize{Optimal Algorithm $(C=1)$}}
\put(30,54){\footnotesize{Trees $1,2 \text{ and } 3$ $(C=1)$}}
\put(30,49){\footnotesize{Trees $1$ and $2$ $(C=6/7)$}}
\put(30,43){\footnotesize{Tree $1$ $(C=3/4)$}}
\end{overpic}
\caption{Average delay performance of the optimal broadcast policy $\pi^{*}$ and the tree-based policy $\pi_{\text{tree}}$ that balances traffic over different subsets of spanning trees.}
\label{fig:601}
\end{figure}

\begin{table*}[ht]
\centering
\begin{tabular}{c | c c c c c | c}
\toprule
& \multicolumn{5}{c|}{tree-based policy $\pi_{\text{tree}}$ over the spanning trees:} & broadcast \\
$\lambda$ & $\mathcal{T}_{1}$ & $\mathcal{T}_{1} \sim \mathcal{T}_{2}$ & $\mathcal{T}_{1} \sim \mathcal{T}_{3}$ & $\mathcal{T}_{1} \sim \mathcal{T}_{4}$ & $\mathcal{T}_{1} \sim \mathcal{T}_{5}$ & policy $\pi^{*}$ \\ [0.5 ex]
\hline
$0.5$ & 12.90 & 12.72 & 13.53 & 16.14 & 16.2 & 11.90\\
$0.9$ & $1.3\times 10^4$ & 176.65 & 106.67 & 34.33 & 28.31 & 12.93\\
$1.9$ & $3.31\times 10^4$ & $1.12\times 10^4$ &$4.92\times 10^3$ & 171.56 & 95.76 & 14.67\\
$2.3$ & $3.63\times 10^4$ & $1.89\times 10^4$ & $1.40\times 10^4$& $1.76\times 10^3$ & 143.68 & 17.35\\
$2.7$ & $3.87\times 10^4$& $2.45\times 10^4$ & $2.03\times 10^4$& $1.1\times 10^4$ & 1551.3 & 20.08\\
$3.1$ & $4.03\times 10^4$ & $2.86\times 10^4$& $2.51 \times 10^4$ & $1.78\times 10^4$ & 9788.1 & 50.39\\
\hline
\end{tabular}
\caption{Average delay performance of the tree-based policy $\pi_{\text{tree}}$ over different subsets of spanning trees and the optimal broadcast policy $\pi^{*}$.}
\label{delay_table}
\end{table*}

\section{Conclusion}\label{sec:conclusion}

We characterize the broadcast capacity of a wireless network under general interference constraints. When the underlying network topology is a DAG, we propose a dynamic algorithm that achieves the wireless broadcast capacity. 
Our novel design, based on packet deficits and the in-order packet delivery constraint, is promising for application to other systems with packet replicas, such as multicasting and caching systems. 
Future work involves the study of cyclic networks, where optimal policies must be sought in the class $\Pi\setminus \Pi^{\text{in-order}}$.

\bibliographystyle{IEEEtran}
\bibliography{MIT_broadcast_bibliography}

\begin{appendix}


\subsection{Proof of Theorem \ref{broadcast_ub}} \label{broadcast_ub_proof}

Fix an $\epsilon >0$. Consider a policy $\pi \in \Pi$ that achieves a broadcast rate of at least $\lambda^* -\epsilon$ defined in~\eqref{bcdef}; this policy $\pi$ exists by the definition of the broadcast capacity $\lambda^{*}$ in Definition~\ref{capacity_def}. Consider any proper cut $U$ of the network $\mathcal{G}$. By definition, there exists a node $i \notin U$. Let $\bm{s}^{\pi}(t) = (s_{e}^{\pi}(t), e\in E)$ be the link-activation vector chosen by policy $\pi$ in slot $t$. The maximum number of packets that can be transmitted across the cut $U$ in slot $t$ is at most $\sum_{e\in E_{U}} c_{e} s_{e}^{\pi}(t)$, which is the total capacity of all activated links across $U$, and the link subset $E_{U}$ is given in~\eqref{eq:604}. The number of distinct packets received by a node $i$ by time $T$ is upper bounded by the total available capacity across the cut $U$ up to time $T$, subject to link-activation decisions of policy $\pi$. That is, we have
\begin{equation} \label{bound_packet}
R_i^{\pi}(T) \leq \sum_{t=1}^{T} \sum_{e\in E_{U}} c_{e} s_{e}^{\pi}(t) = \bm{u}\cdot \sum_{t=1}^{T} \bm{s}^{\pi}(t),
\end{equation}
where we define the vector $\bm{u} = (u_{e}, e\in E)$, $u_{e} = c_{e} 1_{[e\in E_{U}]}$, and $\bm{a}\cdot\bm{b}$ is the inner product of two vectors.\footnote{Note that~\eqref{bound_packet} remains valid if network coding operations are allowed.} Dividing both sides by $T$ yields
\[
 \frac{R_i^{\pi}(T)}{T} \leq \bm{u}\cdot\bigg(\frac{1}{T} \sum_{t=1}^{T}\bm{s}^{\pi}(t)\bigg).
\]
It follows that
\begin{align} 
\lambda^*-\epsilon  &\stackrel{(a)}{\leq} \min_{j\in V} \liminf_{T\to \infty} \frac{R_j^{\pi}(T)}{T}\leq  \liminf_{T\to \infty} \frac{R_i^{\pi}(T)}{T} \nonumber \notag  \\
&\leq  \liminf_{T\to \infty}\bm{u}\cdot\bigg(\frac{1}{T} \sum_{t=1}^{T}\bm{s}^{\pi}(t)\bigg) \label{bound2},
\end{align}
where (a) follows  that $\pi$ is a broadcast policy of rate at least $\lambda^*-\epsilon$. Consider the following useful lemma.

\begin{lemma} \label{conv_hull}
For each proper cut $U$, there exists a vector $\bm{\beta}_U^{\pi} \in \textrm{conv}(\mathcal{S})$ such that 
\[
\liminf_{T\to \infty}  \bm{u}\cdot\bigg(\frac{1}{T} \sum_{t=1}^{T}\bm{s}^{\pi}(t)\bigg) = \bm{u}\cdot \bm{\beta}_U^{\pi}.
\]
\end{lemma}
\begin{IEEEproof}
Consider the sequence $\bm{\zeta}_T^{\pi} = \frac{1}{T}\sum_{t=1}^{T}\bm{s}^{\pi}(t)$ indexed by $T\geq 1$. Since $\bm{s}^{\pi}(t) \in \mathcal{S}$ for all $t\geq 1$, we have $\bm{\zeta}_T^{\pi} \in \conv{\mathcal{S}}$ for all $T\geq 1$. By definition of $\liminf$, there exists a subsequence $\{\bm{u}\cdot \bm{\zeta}_{T_k}^{\pi}\}_{k\geq 1}$ of the sequence $\{\bm{u}\cdot \bm{\zeta}_{T}^{\pi}\}_{T\geq 1}$ such that
\begin{equation} \label{lim}
\lim_{k \to \infty} \bm{u}\cdot \bm{\zeta}_{T_k}^{\pi} = \liminf_{T\to \infty} \bm{u} \cdot \bm{\zeta}_T^{\pi}.
\end{equation} 
Since the set $\conv{\mathcal{S}}$ is closed and bounded, any sequence in $\text{conv}(\mathcal{S})$ has a converging subsequence. Thus, there exists a subsubsequence $\{\bm{\zeta}_{T_{k_i}}^{\pi}\}_{i\geq 1}$ and  $\bm{\beta}_U^{\pi} \in \text{conv}(\mathcal{S})$ such that 
\[
\bm{\zeta}_{T_{k_i}}^{\pi} \to \bm{\beta}_U^{\pi}, \quad \text{as $i\to\infty$.}
\]
It follows that
\begin{align*}
\bm{u} \cdot \bm{\beta}_U^{\pi} &\overset{(a)}{=} \lim_{i \to \infty} \bm{u} \cdot \bm{\zeta}_{T_{k_i}}^{\pi} \overset{(b)}{=} \lim_{k \to \infty} \bm{u}\cdot \bm{\zeta}_{T_k}^{\pi} \overset{(c)}{=} \liminf_{T\to \infty} \bm{u} \cdot \bm{\zeta}_T^{\pi} \\
&= \liminf_{T\to \infty}  \bm{u}\cdot\bigg(\frac{1}{T} \sum_{t=1}^{T}\bm{s}^{\pi}(t)\bigg),
\end{align*}
where (a) uses the fact that 
if $\bm{x}_n \to \bm{x}$ then $\bm{c}\cdot \bm{x}_n \to \bm{c}\cdot \bm{x}$ for $\bm{c}$, $\bm{x}_n$, and $\bm{x} \in \mathbb{R}^l$, $l\geq 1$; (b) follows that if the limit of a sequence $\{z_{n}\}$ exists then all subsequences $\{z_{n_k}\}$ converge and $\lim_{k} z_{n_k}= \lim_n z_n$; (c) follows from Equation~\eqref{lim}.
\end{IEEEproof}

Combining Lemma~\ref{conv_hull} with~\eqref{bound2}, we have that for every proper cut $U$, there exists $\bm{\beta}_U^{\pi} \in \text{conv}(\mathcal{S})$ such that
\begin{equation} \label{bdcutC}
\lambda^*-\epsilon \leq \bm{u} \cdot \bm{\beta}_U^{\pi}.
\end{equation} 
Since~\eqref{bdcutC} holds for all proper cuts $U$, we have
\begin{equation} \label{ub}
 \lambda^* -\epsilon \leq \min_{\text{$U$: a proper cut}} \bm{u} \cdot \bm{\beta}_U^{\pi}.
\end{equation}
Lemma~\ref{conv_hull} shows that $ \beta_{U}^{\pi} \in \conv{\mathcal{S}}$ for any proper cut $U$ and policy $\pi \in \Pi$. Thus, from~\eqref{ub} we get
\begin{equation} \label{conv}
 \lambda^*-\epsilon \leq \sup_{\bm{\beta} \in \conv{\mathcal{S}}} \bigg(\min_{\text{$U$: a proper cut}} \bm{u} \cdot \bm{\beta} \bigg).
\end{equation}
The minimum in~\eqref{conv} is the pointwise minimum of a finite number of linear functions, and is concave and continuous \cite{boyd}. Thus, there exists a vector $\bm{\beta}^* \in\conv{\mathcal{S}}$ that achieves the supremum in the compact set $\conv{\mathcal{S}}$~\cite{bertsekas_convex}. Also,~\eqref{conv} holds for all $\epsilon >0$. We conclude that
\begin{equation} \label{capacity}
\begin{split}
 \lambda^* &\leq \max_{\bm{\beta} \in \text{conv}(\mathcal{S})} \bigg(\min_{\text{$U$: a proper cut}} \bm{u} \cdot \bm{\beta} \bigg) \\
 &= \max_{\bm{\beta} \in \text{conv}(\mathcal{S})} \min_{\text{$U$: a proper cut}} \sum_{e\in E_{U}} c_{e} \beta_{e}.
 \end{split}
\end{equation}

\subsection{Proof of Lemma \ref{in_order}} \label{in_order_proof}
\begin{figure}
\centering
\subfigure[A wired network with a directed cycle $a\to b\to c\to a$.]{
\includegraphics[width=0.17\textwidth]{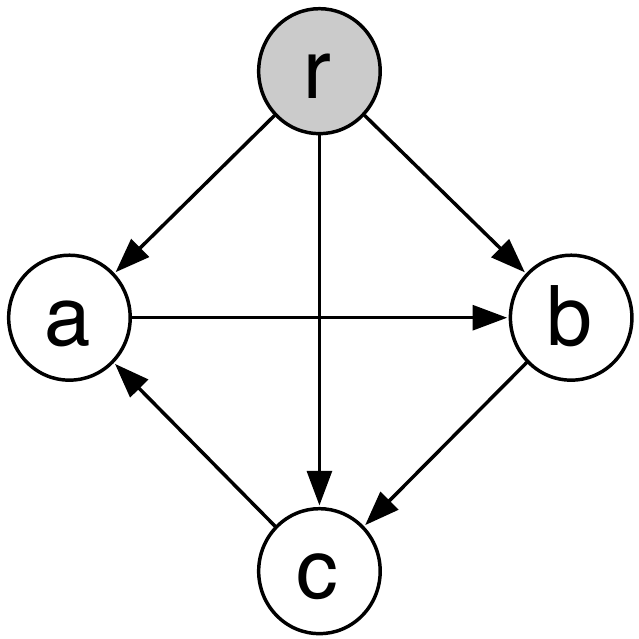}
\label{fig:701}
}
\subfigure[Tree $T_{1}$]{
\includegraphics[width=0.17\textwidth]{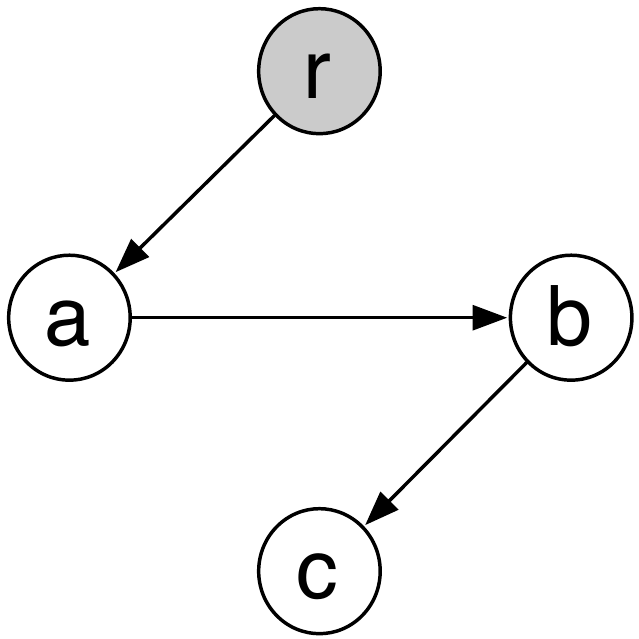}
\label{fig:702}}
\subfigure[Tree $T_{2}$]{
\includegraphics[width=0.17\textwidth]{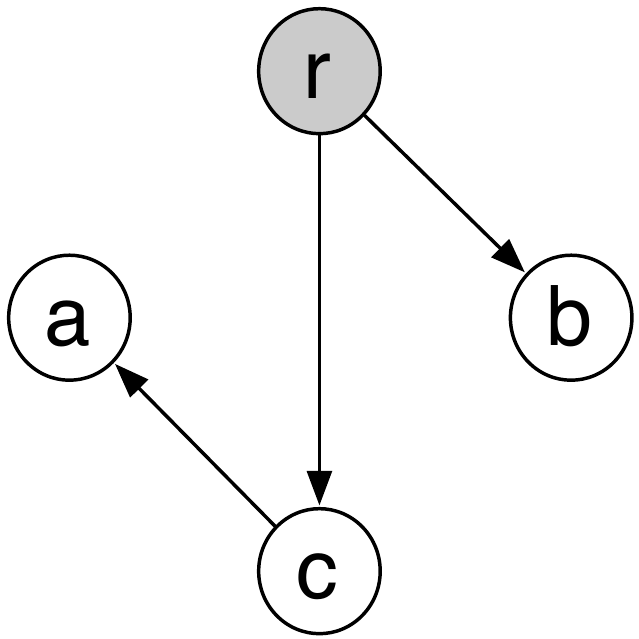}
\label{fig:703}
}
\caption{A cyclic wired network and its two edge-disjoint spanning trees that yield the broadcast capacity $\lambda^{*}=2$.}
\label{trees}
\end{figure}

Consider the cyclic wired network in Fig.~\ref{fig:701}, where all edges have unit capacity and there is no interference constraint. Node $a$ has total incoming capacity equal to two; thus, the broadcast capacity  of the network is upper bounded by $\lambda^{*}\leq 2$. In fact, the network has two edge-disjoint spanning trees as shown in Figures~\ref{fig:702} and~\ref{fig:703}. We can achieve the broadcast capacity $\lambda^{*}=2$ by routing odd and even packets along the trees $T_{1}$ and $T_{2}$, respectively.

Consider a policy $\pi \in \Pi_{\text{in-order}}$ that provides in-order delivery of packets to all network nodes. Let $R_i(t)$ be the number of distinct packets received by node $i$ up to time $t$; node $i$ receives packets $\{1, 2, \ldots, R_{i}(t)\}$ by time $t$ due to in-order packet delivery. Consider the directed cycle $a\to b \to c \to a$ in Fig.~\ref{fig:701}. The necessary condition for all links in the cycle to forward packets in slot $t$ is $R_a(t) > R_b(t) > R_c(t) > R_a(t)$, which is infeasible. Thus, there must exist an idle link in the cycle in every slot. Define the indicator variable $x_{e}(t)=1$ if link $e$ is idle in slot $t$ under policy $\pi$, and $x_{e}(t)=0$ otherwise. Since at least one link in the cycle is idle in every slot, we have
\[
 x_{(a, b)}(t) +  x_{(b, c)}(t) +  x_{(c, a)}(t) \geq 1.
\]
Taking a time average of the above inequality yields
\[
\frac{1}{T} \sum_{t=1}^{T} \big(  x_{(a, b)}(t) +  x_{(b, c)}(t) +  x_{(c, a)}(t) \big) \geq 1.
\]
Taking a $\limsup$ at both sides, we obtain
\begin{multline*}
\sum_{e\in\{(a, b), (b, c), (c, a)\}} \limsup_{T\to\infty} \frac{1}{T}\sum_{t=1}^{T} x_{e}(t) \\
\quad \geq \limsup_{T\to\infty} \sum_{e\in\{(a, b), (b, c), (c, a)\}} \frac{1}{T} \sum_{t=1}^{T} x_{e}(t) \geq 1.
\end{multline*}
The above inequality implies that
\begin{equation} \label{inorder_rate}
\max_{e\in\{(a, b), (b, c), (c, a)\}} \limsup_{T\to\infty} \frac{1}{T}\sum_{t=1}^{T} x_{e}(t) \geq \frac{1}{3}.
\end{equation}
Since the nodes $\{a,b,c\}$ are symmetrically located (i.e., the graph obtained by permuting the nodes $\{a, b, c\}$ is isomorphic to the original graph), without any loss of generality we may assume that the link $e = (a, b)$ attains the maximum in~\eqref{inorder_rate}, i.e.,
\begin{equation} \label{eq:107}
 \limsup_{T\to\infty} \frac{1}{T}\sum_{t=1}^{T} x_{(a, b)}(t) \geq \frac{1}{3}.
\end{equation}
Noting that $x_{e}(t)=1$ if link $e$ is idle in slot $t$ and that node~$b$ receives packets from nodes $r$ and $a$, we can upper bound $R_{b}(T)$ by
\begin{align*}
R_b(T) &\leq \sum_{t=1}^{T} \big(1- x_{(r, b)}(t) +  1-x_{(a, b)}(t)\big) \\
&\leq \sum_{t=1}^{T} \big(2 - x_{(a, b)}(t)\big).
\end{align*}
It follows that
\[
\liminf_{T\to\infty} \frac{R_b(T)}{T} \leq 2 - \limsup_{T\to\infty} \frac{1}{T} \sum_{t=1}^{T} x_{(a, b)}(t) \leq \frac{5}{3},
\]
where the last inequality uses~\eqref{eq:107}. Thus, we have
\[
\min_{i \in V} \liminf_{T\to\infty} \frac{R_{i}(T)}{T} \leq \liminf_{T\to\infty} \frac{R_{b}(T)}{T} \leq \frac{5}{3},
\]
which holds for all policies $\pi\in\Pi_{\text{in-order}}$. Taking the supremum over the policy class $\Pi_{\text{in-order}}$ shows that the broadcast capacity $\lambda^{*}_{\text{in-order}}$ subject to the in-order packet delivery constraint satisfies
\[
\lambda^*_{\text{in-order}} = \sup_{\pi \in \Pi_{\text{in-order}}} \min_{i \in V} \liminf_{T\to\infty} \frac{R_{i}(T)}{T} \leq \frac{5}{3} < 2 = \lambda^*,\footnote{Here, we use the more rigorous definition of a broadcast policy of rate $\lambda$ in~\eqref{eq:703}.}
\]
i.e., the network broadcast capacity is strictly reduced by in-order packet delivery in the cyclic network in Fig.~\ref{fig:701}.

\subsection{Proof of Theorem~\ref{main_theorem}} \label{main_theorem_proof}

We present the proof in three steps. First, using the dynamics of $X_{j}(t)$ in~\eqref{bnd2}, we derive an expression of the one-slot Lyapunov drift using quadratic Lyapunov functions. Second, we design a useful stationary randomized policy that yields near-optimal broadcast throughput; this policy is useful to show that state variables $\{X_j(t)\}$ (treated as virtual queues) under the optimal broadcast policy $\pi^*$ are strongly stable for all arrival rates $\lambda < \lambda^*$.  Third, based on the above analysis, we show that the policy $\pi^*$ is a throughput-optimal broadcast policy for any underlying network graph that is a DAG.

\begin{lemma} \label{algebra}
If we have
\begin{equation} \label{eq:108}
Q(t+1)\leq  (Q(t)-\mu(t))^+ + A(t) 
\end{equation}
where all the variables are non-negative and $(x)^+ = \max\{x,0\}$, then
\[
Q^2(t+1) - Q^2(t) \leq \mu^2(t) + A^2(t) + 2Q(t)(A(t)-\mu(t)).
\]
\end{lemma}
\begin{IEEEproof}
Squaring both sides of~\eqref{eq:108} yields
\begin{align*}
&Q^2(t+1) \\
&\leq  \big((Q(t)-\mu(t))^+\big)^2 + A^2(t) + 2 A(t)(Q(t)-\mu(t))^+\\
&\leq  (Q(t)-\mu(t))^2 + A^2(t) + 2 A(t)Q(t),
\end{align*}
where we use the fact that $x^2 \geq {(x^+)}^2$, $Q(t) \geq 0$, and $\mu(t) \geq 0$. Rearranging the above inequality finishes the proof.
\end{IEEEproof}
Applying Lemma~\ref{algebra} to the dynamics~\eqref{bnd2} of $X_{j}(t)$ yields, for each node $j\neq r$,
\begin{multline} \label{eq:109}
X_j^2(t+1) - X_j^2(t) \\
\leq  B(t) + 2 X_j(t) \big(\sum_{m\in V}\mu_{mi_{t}^*}(t)-\sum_{k\in V} \mu_{kj}(t)\big),
\end{multline}
where $B(t)\leq \mu^2_{\max}+ \max\{a^2(t),\mu^2_{\max}\} \leq  (a^2(t) + 2\mu^2_{\max})$, $a(t)$ is the number of exogenous packet arrivals in a slot, and $\mu_{\max} \triangleq \max_{e\in E} c_e$ is the maximum capacity of the links.
We assume the arrival process $a(t)$ has bounded second moments; thus, there exists a finite constant $B>0$ such that $\mathbb{E}[B(t)] \leq \mathbb{E}\big(a^2(t)\big) + 2\mu^2_{\max} < B$.

We define the quadratic Lyapunov function $L(\bm{X}(t)) = \sum_{j\neq r} X_j^2(t)$. From~\eqref{eq:109}, the one-slot Lyapunov drift $\Delta(\bm{X}(t))$ satisfies
\begin{align} \label{drift2}
&\Delta(\bm{X}(t)) \triangleq  \mathbb{E}[L(\bm{X}(t+1) - L(\bm{X}(t)) \mid \bm{X}(t)] \notag \\
&= \mathbb{E}\big[\sum_{j\neq r} \big(X_j^2(t+1) - X_j^2(t) \big) \mid \bm{X}(t)\big] \notag \\
&\leq B|V| +2  \sum_{j\neq r} X_{j}(t) \mathbb{E}\big[\sum_{m\in V}\mu_{mi_{t}^*}(t)-\sum_{k\in V} \mu_{kj}(t) \mid \bm{X}(t)\big] \notag \\
&= B|V| - 2 \sum_{(i,j)\in E} \mathbb{E}[\mu_{ij}(t)\mid\bm{X}(t)] \big( X_j(t) - \sum_{k\in K_{j}(t)} X_k(t) \big) \notag \\
&= B|V|- 2 \sum_{(i,j)\in E} \mathbb{E}[\mu_{ij}(t)\mid \bm{X}(t)] \, W_{ij}(t),
\end{align}
where $K_{j}(t)$ and $W_{ij}(t)$ are defined in~\eqref{eq:110} and~\eqref{eq:111}, respectively. To emphasize that the evaluation of the inequality~\eqref{drift2} depends on a control policy $\pi\in\Pi^{*}$, we rewrite~\eqref{drift2} as
\begin{equation} \label{eq:112}
\Delta^{\pi}(\bm{X}(t)) \leq B|V|- 2 \sum_{(i,j)\in E} \mathbb{E}[\mu^{\pi}_{ij}(t)\mid \bm{X}(t)] \, W_{ij}(t).
\end{equation}
Our optimal broadcast policy $\pi^{*}$ is chosen to minimize the drift on the right-hand side of~\eqref{eq:112} among all policies in $\Pi^{*}$.

We construct a useful randomized policy $\pi^{\text{RAND}} \in\Pi^{*}$. Let $\bm{\beta}^*\in\conv{\mathcal{S}}$ be the vector that attains the outer bound on the broadcast capacity $\lambda^{*}$ in Theorem~\ref{broadcast_ub}, i.e.,
\[
\bm{\beta}^* \in \arg \max_{\bm{\beta} \in \conv{\mathcal{S}}} \min_{\text{$U$: a proper cut}} \sum_{e\in E_{U}} c_{e} \beta_{e}.
\]
From the Caratheodory's theorem~\cite{bertsekas_convex}, there exist at most $(|E|+1)$ link-activation vectors $\bm{s}_l\in \mathcal{S}$ and the associated nonnegative scalars $\{p_l\}$ with $\sum_{l=1}^{|E|+1}p_l=1$, such that 
\begin{equation} \label{beta_star}
\bm{\beta}^*= \sum_{l=1}^{|E|+1} p_l \bm{s}_l.
\end{equation}
From Theorem~\ref{broadcast_ub}, we get
\begin{equation} \label{bc_bound}
\lambda^* \leq \min_{\text{$U$: a proper cut}} \sum_{e\in E_{U}} c_{e} \beta_{e}^{*}.
\end{equation}
Suppose that the exogenous packet arrival rate $\lambda$ is strictly less than the broadcast capacity $\lambda^*$. There exists an $\epsilon >0$ such that $\lambda +\epsilon \leq \lambda^{*}$. From~\eqref{bc_bound}, we have
\begin{equation} \label{eq:113}
\lambda+\epsilon \leq \min_{\text{$U$: a proper cut}} \sum_{e\in E_{U}} c_{e} \beta_{e}^{*}.
\end{equation}
For any network node $v\neq r$, consider the proper cut $U_{v} = V\setminus \{v\}$. We have, from~\eqref{eq:113}, that
\begin{equation} \label{capacity_exceeding}
\lambda + \epsilon \leq  \sum_{e\in E_{U_{v}}} c_{e} \beta_{e}^{*}, \ \forall v\neq r.
\end{equation}
Since the underlying network topology $\mathcal{G}=(V, E)$ is a DAG, we have a topological ordering  of the network nodes so that: (i) all links in $E$ are directed from left to right~\cite{algorithms}; (ii) we can label the network nodes serially as $\{v_{1}, \ldots, v_{|V|}\}$, where $v_{1}$ is the source node with no in-neighbors and $v_{|V|}$ has no outgoing neighbors. From~\eqref{capacity_exceeding}, we define $q_{l}\in[0, 1]$ for each node $v_{l}$ such that 
\begin{equation} \label{q_prob}
q_{l}\, \sum_{e\in E_{U_{v_{l}}}} c_{e} \beta_{e}^{*} = \lambda + \epsilon \frac{l}{|V|},\  l=2, \ldots , |V|.
\end{equation}
Consider the randomized broadcast policy $\pi^{\text{RAND}} \in \Pi^{*}$ working as follows: (i) it selects the feasible link-activation vector $\bm{s}(t) = \bm{s}_{l}$ with probability $p_{l}$ in~\eqref{beta_star}, $l=1,2, \ldots, |E|+1$, in every slot $t$; (ii) for each incoming link $e = (\cdot, v_{l})$ of node $v_{l}$ such that $s_{e}(t)=1$, the link $e$ is activated independently with probability $q_{l}$; (iii) activated links are used to forward packets, subject to the constraints that define the policy class $\Pi^{*}$ (i.e., in-order packet delivery and that a network node is only allowed to receive packets that have been received by all of its in-neighbors). Note that this randomized policy does not use the state variables $\bm{X}(t)$.  Since each network node $j$ is relabeled as $v_{l}$ for some $l$, from~\eqref{q_prob} we have, for each node $j\neq r$, the total expected incoming transmission rate satisfies
\begin{align} 
\sum_{i: (i, j)\in E}\mathbb{E}[\mu^{\pi^{\text{RAND}}}_{ij}(t)\mid\bm{X}(t)] &=\sum_{i: (i,j)\in E} \mathbb{E}[\mu^{\pi^{\text{RAND}}}_{ij}(t)]  \notag \\
&= q_{l}\, \sum_{e\in E_{U_{v_{l}}}} c_{e} \beta_{e}^{*} \notag \\
&=\lambda + \epsilon \frac{l}{|V|}. \label{rate_comp1}
\end{align}
Equation~\eqref{rate_comp1} shows that the randomized policy $\pi^{\text{RAND}}$ provides each network node $j\neq r$ with the total time-average incoming capacity strictly larger than the packet arrival rate $\lambda$ via proper random link activations. According to the abuse of notation in~\eqref{bnd2}, at the source node $r$ we have
\begin{equation} \label{rate_comp2}
\sum_{i:(i,r)\in E} \mathbb{E}[\mu^{\pi^{\text{RAND}}}_{ir}(t)\mid\bm{X}(t)] = \mathbb{E}[\sum_{i:(i,r)\in E} \mu^{\pi^{\text{RAND}}}_{ir}(t)] = \lambda.
\end{equation}
From~\eqref{rate_comp1} and~\eqref{rate_comp2}, if node $i$ lies before $j$ in the aforementioned topological ordering, i.e., $i = v_{l_{i}} < v_{l_{j}} = j$ for some $l_{i} < l_{j}$, then
\begin{align} 
&\sum_{k:(k,i)\in E}\mathbb{E}[\mu^{\pi^{\text{RAND}}}_{ki}(t)\mid\bm{X}(t)]- \sum_{k:(k,j)\in E}\mathbb{E}[\mu^{\pi^{\text{RAND}}}_{kj}(t)\mid\bm{X}(t)]  \notag \\
&\leq -\frac{\epsilon}{|V|}. \label{rate_comparison_final}
\end{align}
The drift inequality~\eqref{drift2} holds for any policy $\pi \in \Pi^*$. Our broadcast policy $\pi^{*}$ observes the system states $\bm{X}(t)$ and seek to minimize the drift in every slot. Comparing the actions taken by the policy $\pi^{*}$ with those by the randomized policy $\pi^{\text{RAND}}$ in slot $t$ in~\eqref{drift2}, we get
\begin{align}
&\Delta^{\pi^*}(\bm{X}(t)) \leq B|V|- 2 \sum_{(i,j)\in E}\mathbb{E}\big[\mu^{\pi^{*}}_{ij}(t) \mid\bm{X}(t)] W_{ij}(t) \notag \\
&\leq B|V|- 2 \sum_{(i,j)\in E}\mathbb{E}\big[\mu^{\pi^{\text{RAND}}}_{ij}(t) \mid\bm{X}(t)] W_{ij}(t) \notag \\
&=  B|V| +2 \sum_{j\neq r} X_j(t) \bigg(\sum_{m\in V}\mathbb{E}\big[\mu^{\pi^{\text{RAND}}}_{mi_{t}^*}(t)\mid\bm{X}(t)\big] \notag \\
&\quad -\sum_{k\in V} \mathbb{E}\big[\mu^{\pi^{\text{RAND}}}_{kj}(t)|\bm{X}(t)\big]\bigg) \notag \\
&\leq B|V| - \frac{2\epsilon}{|V|} \sum_{j\neq r} X_{j}(t). \label{eq:114}
\end{align}
Note that $i_{t}^*= \arg \min_{i\in \text{In}(j)} Q_{ij}(t)$ for a given node $j$. Since node $i_{t}^{*}$ is an in-neighbor of node $j$, $i_{t}^{*}$ must lie before $j$ in any topological ordering of the DAG. Hence, the last inequality of~\eqref{eq:114} follows~\eqref{rate_comparison_final}. Taking expectation in~\eqref{eq:114} with respect to $\bm{X}(t)$, we have
\[
\mathbb{E}\big[L(\bm{X}(t+1))\big]-\mathbb{E}\big[L(\bm{X}(t))\big] \leq B|V| -\frac{2\epsilon}{|V|}\mathbb{E}||\bm{X}(t)||_1,
\]
where $||\cdot ||_1$ is the $\ell_1$-norm of a vector. Summing the above over $t=0, 1,2,\ldots T-1$ yields
\[
\mathbb{E}\big[L(\bm{X}(T))\big]-\mathbb{E}\big[L(\bm{X}(0))\big] \leq B|V|T -\frac{2\epsilon}{|V|}\sum_{t=0}^{T-1}\mathbb{E}||\bm{X}(t)||_1.
\]
Dividing the above by $2T\epsilon/|V|$ and using $L(\bm{X}(t))\geq 0$, we have 
\begin{eqnarray*}
\frac{1}{T}\sum_{t=0}^{T-1}\mathbb{E}||\bm{X}(t)||_1 \leq \frac{B|V|^2}{2\epsilon} + \frac{|V|\,\mathbb{E}[L(\bm{X}(0))]}{2T\epsilon}
\end{eqnarray*}
Taking a $\limsup$ of both sides yields
\begin{eqnarray} \label{strong_stability}
\limsup_{T \to \infty}\frac{1}{T}\sum_{t=0}^{T-1} \sum_{j\neq r} \mathbb{E}[X_{j}(t)]  \leq \frac{B|V|^2}{2\epsilon}
\end{eqnarray} 
and all queues $X_{j}(t)$ are strongly stable.

Next, we show that the strong stability of the virtual queues $X_{j}(t)$ implies that the policy $\pi^*$ achieves the broadcast capacity $\lambda^{*}$, i.e., for all arrival rates $\lambda < \lambda^*$, we have 
\[
 \lim_{T\to \infty} \frac{R_j(T)}{T} = \lambda,  \ \forall j.
\]
Equation~\eqref{bnd2} shows that the virtual queues $X_{j}(t)$ have bounded departures (due to the finite link capacities). Thus, strong stability of $X_{j}(t)$ implies that all virtual queues $X_{j}(t)$ are rate stable~\cite[Theorem~$2.8$]{neely2010stochastic}, i.e., $\lim_{T\to \infty} X_j(T)/T = 0$ for all $j$. It follows that
\begin{equation} \label{rate_stability_of_X}
\lim_{T\to \infty} \frac{\sum_{j\neq r}X_j(T)}{T} = 0.
\end{equation}
Now consider any node $j\neq r$ in the network. Let us construct a simple path $\sigma (r = u_n \to u_{n-1} \ldots \to u_1 = j)$ from the source node $r$ to the node $j$ by running the following algorithm on the underlying graph $\mathcal{G}(V,E)$. 
 \begin{algorithm} 
\caption{$r\to j$ Path Construction Algorithm}
\begin{algorithmic}[1] 
 \REQUIRE Graph $\mathcal{G}(V,E)$, node $j\in V$
 \STATE $i \gets 1$
 \STATE $u_i\gets j$
 \WHILE{$u_i \neq r$} 
 \STATE $u_{i+1} \gets \arg \min_{k\in\text{In}(u_{i})} Q_{ku_i}(t)$; ties are broken arbitrarily.
 \STATE $i \gets i+1$
 \ENDWHILE
 \end{algorithmic}
 \end{algorithm} 
 
This algorithm chooses the parent of a node $u$ in the path $\sigma$ as the one that has the least relative packet deficit as compared to $u$. Since the underlying graph $\mathcal{G}(V,E)$ is a connected DAG (i.e., there is a path from the source to every other node in the network), the above path construction algorithm always terminates with a path $\sigma(r\to j)$. The number of distinct packets received by node $j$ up to time $T$ can be written as a telescoping sum of relative packet deficits along the path $\sigma$, i.e.,
\begin{align}
R_j(T) &= R_{u_1}(T) \notag \\
&= \sum_{i=1}^{n-1}\big(R_{u_i}(T)-R_{u_{i+1}}(T)\big) +R_{u_n}(T)  \notag \\
&= -\sum_{i=1}^{n-1} X_{u_i}(T) + R_r(T) \notag \\
&= -\sum_{i=1}^{n-1} X_{u_i}(T) + \sum_{t=0}^{T-1} a(t), \label{eq:116}
\end{align}
where the third equality follows the observation that (see~\eqref{eq:115})
\[
X_{u_{i}}(T) = Q_{u_{i+1}u_{i}}(T) = R_{u_{i+1}}(T) - R_{u_{i}}(T).
\]
Using $\sum_{i=1}^{n-1} X_{u_{i}}(t) \leq \sum_{j\neq r} X_{j}(t)$,~\eqref{eq:116}, and that $X_{j}(t)$ are nonnegative, we have, for each node $j$,
\[
\frac{1}{T}\sum_{t=0}^{T-1} a(t) -\frac{1}{T}\sum_{j\neq r} X_{j}(T) \leq \frac{1}{T} R_j(T) \leq \frac{1}{T}\sum_{t=0}^{T-1} a(t).
\]
Taking a limiting time average and the strong law of large numbers for the arrival process, we have
\[
 \lim_{T\to \infty} \frac{R_j(T)}{T} = \lambda, \, \forall j.
\]
This concludes the proof.

\subsection{Proof of Lemma~\ref{lem:701}} \label{pf:701}
We regard the DAG $G$ as a wired network in which all links can be activated simultaneously. Theorem~\ref{main_theorem} and~\eqref{eq:603} show that the broadcast capacity of the wired network $G$ is
\begin{align}
\lambda^{*} = \lambda_{\text{DAG}} = \min_{\text{$U$: a proper cut}}\, \sum_{e\in E_{U}} c_{e}  &= \min_{\{U_v, v\neq r\} }\, \sum_{e\in E_{U_{v}}} c_{e} \notag \\
&= \min_{v\in V\setminus \{r\}} d_{\text{in}}(v), \label{eq:605}
\end{align}
where $U_{v} = V \setminus \{v\}$ is the proper cut that separates node $v$ from the network, $E_{U_{v}}$ is the set of incoming links of node $v$, and the last equality follows that the maximizer in~\eqref{eq:603} is the all-one vector $\bm{\beta} = \bm{1}$ and that all links have unity capacity. The Edmond's theorem~\cite{edmonds} states that the maximum number of disjoint spanning trees in the directed graph $G$ is
\begin{equation} \label{ed1}
k^* =\min_{\text{$U$: a proper cut}} \sum_{e\in E_{U}} c_e.
\end{equation}
Combining~\eqref{eq:605} and~\eqref{ed1} completes the proof.

\end{appendix}

\end{document}